\documentclass[sigconf]{acmart}

\pdfoutput=1 
\usepackage{siunitx}
\usepackage{balance}
\usepackage{xcolor}
\usepackage{listings}
\usepackage{multirow} 
\usepackage{mdframed} 

\definecolor{lightyellow}{rgb}{1,1,0.8}
\usepackage{algorithm}
\usepackage{algpseudocode}

\lstdefinelanguage{json}{
    basicstyle=\ttfamily\small,
    keywordstyle=\color{blue},
    stringstyle=\color{black},
    showstringspaces=false,  
    morestring=[b]",
    morestring=[d]'
}

\mdfdefinestyle{mystyle}{
  backgroundcolor=lightyellow,
  linewidth=0pt, 
  innerleftmargin=5pt,
  innerrightmargin=10pt,
  innertopmargin=10pt,
  innerbottommargin=5pt
}

\usepackage{mdframed}
\usepackage{setspace}
\definecolor{lightbrown}{RGB}{255,249,241}
\newenvironment{promptbox}
    {\begin{mdframed}[backgroundcolor=lightbrown,linecolor=white]
        \begin{small}
            \begin{spacing}{0.9}}
    {\end{spacing}
    \end{small}
    \end{mdframed}
    }

\AtBeginDocument{%
  \providecommand\BibTeX{{%
    \normalfont B\kern-0.5em{\scshape i\kern-0.25em b}\kern-0.8em\TeX}}}

\setcopyright{acmcopyright}
\copyrightyear{2018}
\acmYear{2018}
\acmDOI{XXXXXXX.XXXXXXX}

\acmConference[Conference acronym 'XX]{Make sure to enter the correct
  conference title from your rights confirmation emai}{June 03--05,
  2018}{Woodstock, NY}
\acmBooktitle{Woodstock '18: ACM Symposium on Neural Gaze Detection,
 June 03--05, 2018, Woodstock, NY} 
\acmPrice{15.00}
\acmISBN{978-1-4503-XXXX-X/18/06}
\usepackage{comment}

\usepackage{xspace}
\newcommand{\system}{Tutorly\xspace} 
\newcommand{\systems}{Tutorly's\xspace}
\newcommand{\framework}{\textit{CogApp}\xspace} 

\begin{document}

\title{\system: Turning Programming Videos Into Apprenticeship Learning Environments with LLMs}


 \author{Wengxi Li}
 \affiliation{%
  \institution{University of Michigan}
   \country{USA}
 }
 \email{wengxili@umich.edu}

 \author{Roy Pea}
 \affiliation{%
  \institution{Stanford University}
   \country{USA}
 }
 \email{roypea@stanford.edu}

  \author{Nick Haber}
 \affiliation{%
 \institution{Stanford University}
   \country{USA}
 }
 \email{nhaber@stanford.edu}

\author{Hari Subramonyam}
 \affiliation{%
  \institution{Stanford University}
   \country{USA}
 }
 \email{harihars@stanford.edu}

\renewcommand{\shortauthors}{Subramonyam, et al.}

\begin{abstract}
Online programming videos, including tutorials and streamcasts, are widely popular and contain a wealth of expert knowledge. However, effectively utilizing these resources to achieve targeted learning goals can be challenging. Unlike direct tutoring, video content lacks tailored guidance based on individual learning paces, personalized feedback, and interactive engagement necessary for support and monitoring. Our work transforms programming videos into one-on-one tutoring experiences using the cognitive apprenticeship framework. \system, developed as a JupyterLab Plugin, allows learners to (1) set personalized learning goals, (2) engage in learning-by-doing through a conversational LLM-based mentor agent, (3) receive guidance and feedback based on a student model that steers the mentor moves. In a within-subject study with 16 participants learning exploratory data analysis from a streamcast, \system significantly improved their performance from 61.9\% to 76.6\% based on a post-test questionnaire. \system demonstrates the potential for enhancing programming video learning experiences with LLM and learner modeling.

\end{abstract}

\begin{teaserfigure}
  \centering
   \includegraphics[width=\textwidth]{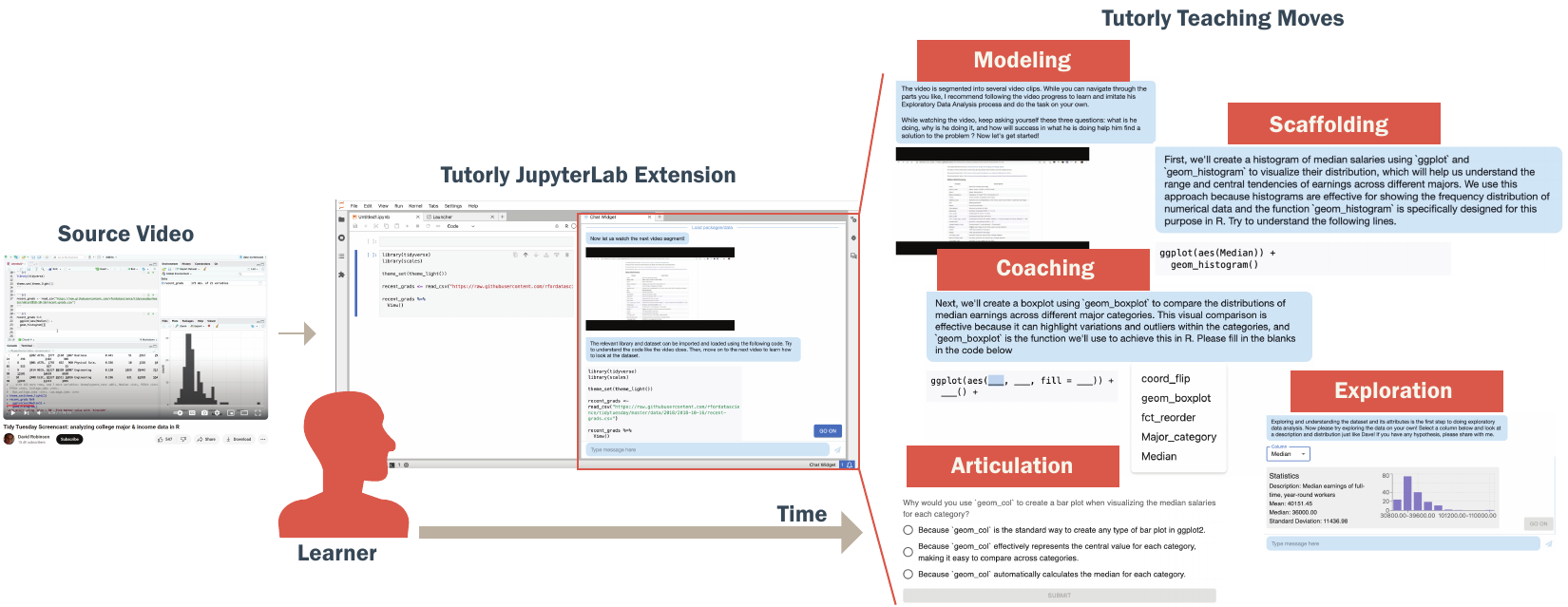}
 \caption{Overview of \system: As the learner engages with a programming video tutorial using our JupyterLab Extension, \system guides the learners through rich multi-modal conversations that correspond to different moves in the Cognitive Apprenticeship Framework. }
  \label{fig:teaser}
  \Description[]{}
\end{teaserfigure}

\maketitle

\section{Introduction}\

Video has emerged as a preferred medium for learning programming. According to a recent survey by Stack Overflow, developers who learned through ``How-to-videos'' and ``Video-based Online Courses'' accounted for 60.14\% and 49.41\%, respectively. Video learning has several advantages, including convenient and asynchronous access~\cite{Mukhtar2020}, flexibility to learn at different paces and focus on selective content~\cite{akhter2021highlighting}, and several online content such as on YouTube are free or available at relatively low costs~\cite{Geith2008ACCESSTE}.  Regardless, the most effective learning paradigm for programming videos is \textit{learning by doing}~\cite{Brown1989SituatedCA}. However, this is easier said than done. Videos and supplement materials are often static and spread out across different content formats and tooling (e.g., videos, slides, or GitHub repositories).  Further, practicing coding tasks while watching videos is a difficult skill to learn, requiring clear instructions and immediate feedback~\cite{deliberate-practice}.  While sometimes students are able to interact with teachers who record videos (e.g., online office hours), the process is time-consuming and burdensome\cite{Ong2023EnhancingTI}.

Simulating instructor-led, supportive practice based on video content could enhance the learning experience and efficacy, enabling students to master key concepts presented in the videos more effectively. We propose that, given the generative power of large language models (LLMs), they offer the opportunity to assume the role of a teacher and provide guidance and immediate feedback for students. However, operationalizing LLMs as tutors for video learning requires addressing a number of current challenges. \textbf{First, current LLMs suffer from a verbosity bias,} which means that they favor lengthy answers even if they are of poorer quality and harder for people to understand compared to shorter answers~\cite{zhao2023survey}. \textbf{Second, due to the open-ended nature of LLM text generation, providing step-by-step and targeted guidance is challenging.} It is usually hard for LLMs to teach students step-by-step what they need to learn in a video, just through prompts~\cite{lightman2023lets}. In contrast, students benefit significantly from a one-to-one instructor who can choose and utilize pedagogy that suits specific students~\cite {vanlehn2011relative}. As a result, the instructor needs to summarize the knowledge and tasks in the video, teach the knowledge in concise sentences, and guide the students to work on their tasks~\cite{Narasimhan2022TLDWSI}. Currently, LLM in a fully open-ended setting is not the answer to applying these strategies and sometimes can even be a significant burden for students~\cite{ren2024human}.

To address the challenges of utilizing LLMs as an intelligent \textit{instructor}, we adopt cognitive strategies that play a crucial role in knowledge acquisition and skill learning. Specifically,  we draw on the \textit{Cognitive Apprenticeship (\framework)} framework from the learning sciences as the foundation for our approach~\cite{collins1991cognitive}. The \framework framework advocates six teaching methods (moves): \textit{Modeling, Coaching, Scaffolding, Articulation, Reflection, and Exploration} that make the instructor's thinking visible to learners and the student's thinking visible to the instructor. For example, a teacher might use \textit{Modeling} to demonstrate the process of solving mathematics problems by having students bring difficult new problems for her to solve in class. \textit{Scaffolding} can take diverse forms, such as suggestions to facilitate writing or physical support in downhill skiing. \textit{Articulation} involves encouraging students to articulate their thoughts as they carry out their problem-solving. The \framework framework also advocates for the selection of teaching methods based on accurate assessments of a student's current skill level and then identifying an intermediate step of suitable difficulty to facilitate the target learning activity~\cite{Pea2004TheSA}.

Leveraging these insights, we introduce \system, an LLM-powered apprenticeship learning environment for programming videos. \system is implemented as a JupyterLab~\cite{ProjectJupyter} extension that supports conversational interactions for scaffolding and guidance in learning from videos. We developed a prompt generation pipeline that segments videos into distinct learning goals and, for each learning goal, selects the appropriate move from the \framework framework by using an internal student model. Rather than allowing LLMs to autonomously choose the method for generating the conversation --- a process that can lead to unpredictable outcomes due to the diversity of videos and student models --- we adopt a more structured approach. Initially, we extract both procedural and conceptual knowledge from video segments that align with distinct learning goals. Subsequently, we select suitable methods for each type of knowledge, informed by insights from student models. Finally, leveraging these methods, along with corresponding actions and interactions, we craft targeted messages. This systematic process ensures a more controlled and relevant message generation tailored to the specific learning needs of students. As shown in Figure~\ref{fig:teaser}, the system progressively generates instructions for students to learn how to visualize the data, starting with MENTORING, then COACHING, and at a later point, SCAFFOLDING and REFLECTION. 

We conducted a technical analysis of \system prompting and a user study of \system's end-to-end effectiveness as a system for helping students while watching videos. We found the guidance generated by our prompting pipeline is more concise, consistent with the learning goal and mentor move. Based on our user study, participants demonstrated better learning outcomes with \system. Our approach for tightly integrating \framework framework with LLMs \textbf{allows for targeted instruction} on knowledge from videos and \textbf{alleviates teaching constraints} through open-ended guidance generation. In summary, \system highlights the potential of generative AI as a virtual instructor for online programming learning, combining pedagogy research on cognitive apprenticeship theory with research on the generative capabilities of LLMs. We contribute:
\begin{enumerate}
  \item \system: an interactive conversational system as an extension in JupyterLab. Given the video, \system generates student-specific guidance and lets students use the interactions to practice knowledge according to the learning goal.
  \item \textit{\framework-based Prompting}: A domain-specific language for organizing cognitive apprenticeship methods, allowing LLMs to generate direction and interactions consistently and stably. It also supports teachers to personalize the combination of methods, interactions, and learning goals.
  \item An evaluation of \textit{\framework prompting pipeline} and a user study of \system with $N = 16$ participants. Our studies highlight \systems effectiveness in applying cognitive apprenticing pedagogy compared to the status quo of teaching the same material.
\end{enumerate}
\section{Related Work}
The capabilities of LLMs offer novel perspectives to assist students in learning programming while watching tutorial videos. Here, we synthesize relevant literature on learning by watching videos, cognitive apprenticeship framework, and LLMs in intelligent tutoring systems (ITS), which is the basis for the design of \system.

\subsection{Video-based Learning}
Video-based learning is a pedagogical approach in which learners acquire knowledge and skills through video content, leveraging multimedia content to enhance student engagement and instructional delivery. In the context of programming and software development education, several studies have investigated the efficacy of video tutorials~\cite{Giannakos2013ExploringTV, Poquet2018VideoAL, Sabli2020VideoBasedL, Schwartz2014s}. Online learning platforms like MOOCs and video platforms like YouTube can deliver tutorial videos in a scalable way. However, a significant challenge is this form of learning does not encourage active engagement. Videos provide few opportunities for the viewer to practice and develop their skills unless they are intrinsically motivated~\cite{10.1145/3274319, Chen2020TowardsSP, Kim2017APA}. Besides, many learning theories emphasize that in order to become responsible and autonomous learners, students need to take control of their learning~\cite{10.1145/3159450.3159506}, express their `epistemic agency'~\cite{Scardamalia2002CollectiveCR, Ko2019OpeningUC, Stroupe2014ExaminingCS}, and participate in creativity-boosting activities alongside explicit instructions on computer programming~\cite{10.1007/s10639-023-11629-4, Alario-Hoyos2016}. Empirical studies also show that the more students engage with the learning content, such as through extensive practice and making mistakes, the easier it is for them to retrieve it~\cite{McGlynn2005TeachingM, alario2016interactive, Hackathorn2011LearningBD, metcalfe2024learning}.

Multiple research has delved into enhancing video-based learning through technology and human-centered design. For example, Guo et al. ~\cite{Guo2014HowVP} highlight how video production quality and presentation style significantly affect student engagement in online courses, emphasizing the need for well-designed educational videos to maintain attention and facilitate learning. ToolScape~\cite{Kim2014CrowdsourcingSI} explores the intricacies of extracting step-by-step information from existing how-to videos, suggesting that enhanced metadata and interactive video components can significantly improve video content's usability and educational value. This is complemented by investigating in-video dropouts and interaction peaks in online lecture videos~\cite{Kim2014UnderstandingID}, which offers insights into student engagement patterns and learning behaviors. VT-Revolution~\cite{Bao2019VTRevolutionIP}, LV4LP~\cite{Lin2022TeachingPB}, and ITSS~\cite{Ouh2022ITSSIW} also incorporate interactive workflow at the time of video creation. Codemotion~\cite{Khandwala2018CodemotionET} and psc2code~\cite{Bao2020EnhancingDI} demonstrated the extraction of executable code from programming tutorial videos, which facilitates active learning by allowing learners to interact directly with the code discussed in the videos. 

To maximize student engagement, interactive programming video tutorial authoring and watching systems have been explored \cite{Guo2015CodeopticonRO, Chen2020TowardsSP, cao2022videosticker}. Zhao et al.~\cite{Zhao2020VideoQA} emphasized the benefits of video question-answering systems for screencast tutorials, which support learners in navigating through complex software learning processes by providing direct answers to specific queries raised during video playback. Soloist~\cite{Wang2021SoloistGM} uses audio processing to generate mixed-initiative tutorials from existing guitar instructional videos, showcasing the potential for automated educational content creation. Moreover, FlowMatic~\cite{Zhang2020FlowMaticAI} points toward the future of educational environments on immersive authoring tools for interactive scenes in virtual reality, where learners can manipulate and experiment with the video content, enhancing the depth of interaction and engagement. To align the interactivity features with learning goals, some works enable users to find segments where a certain sub-topic or keyword is presented and allows the learner to click on them and to jump there directly~\cite{Mahapatra2018VideoKenAV, Yadav2016ViZigAP, Kravvaris2018AutomaticPO, Husain2019MultimodalFO}. Our work builds on these prior approaches and leverages the video content to create interactive learning experiences aligned with learning goals.

\subsection{Cognitive Apprenticeship in Computer Science Education}
Knowledge integration, in which learners acquire, modify, and store learned information with what they already know, is central to a successful learning experience. In domains such as programming, it involves combining multiple \textit{representations} into existing knowledge and fostering a unified understanding of a complex domain~\cite{10.1007/BF0221405}. To facilitate such understanding, work in instructional design has identified patterns to help learners, such as eliciting, adding, distinguishing, and sorting our ideas through analysis and reflection~\cite{Linn_Bell_Davis_2013}. A computational cognitive apprenticeship framework for embodying these patterns should focus not only on ``how to do,'' but ``how to think''~\cite{8661523}. This approach is in line with cognitive load theory, which suggests that learners manage intrinsic, extraneous, and germane cognitive loads during the learning process~\cite{Margulieux2016EmployingSI}. 

In the realm of computer programming education, cognitive apprenticeship has proven to be effective in enhancing learning outcomes. Studies have demonstrated the advantages of incorporating subgoals in programming education, as breaking down complex tasks into smaller subgoals can decrease cognitive load and improve comprehension~\cite{Margulieux2016EmployingSI}. Moreover, providing explicit programming strategies to adolescents has been shown to be beneficial, as it equips them with structured problem-solving approaches and encourages self-regulation throughout the programming process~\cite{Ko2019TeachingEP}. By engaging in authentic tasks, novices can not only acquire the explicit knowledge needed for a task but also the implicit knowledge that experts possess~\cite{Loksa2016TheRO}. Cognitive apprenticeship framework highlights the significance of learning through guided experiences, where novices collaborate with experts to enhance their skills and establish habit in self-regulation~\cite{collins1991cognitive, Casey1996IncorporatingCA}.

The utilization of technology, such as multimedia content creation tools, can facilitate collaborative learning experiences within the cognitive apprenticeship framework. For instance, studies have looked at anchored instruction~\cite{barron2014doing} that integrates computation within disciplinary engineering practices coupled with scaffolding approaches~\cite{10.1145/2534971, 10.1007/s12528-020-09267-7, 10.1016/j.chb.2016.03.025}. Recent work has embedded the methods in computational notebooks to help students learn an unsupervised ML algorithm~\cite{10.1002/cae.22580}. Our work implements the computational cognitive apprenticeship framework with subgoal setting, explicit programming instruction, and collaborative multimedia tools within computational notebook environment to facilitate student practice activities and integrate knowledge.

\subsection{Conversational Intelligent Tutoring Systems}
Conversational Intelligent Tutoring Systems (CITS) have emerged as a promising approach to personalized and interactive learning experiences, which guide students through scaffolding practice problems and provide correctness feedback, next-step hints, and adaptive feedback messages. Latham et al.~\cite{Latham2011OscarAI} introduce Oscar, an Intelligent Adaptive Conversational Agent Tutoring System that incorporates human-like natural language interfaces to support constructivist learning styles. My Science Tutor~\cite{Ward2013MyST} is a conversational multimedia virtual tutor designed to provide individualized and adaptive instruction akin to human tutors, emphasizing the role of ITS in enhancing learning achievement. Furthermore, Ji et al.~\cite{Ji2022ConversationalIT} investigate student and tutor perceptions of conversational ITS in online learning programs, aiming to understand the potential impact of chatbot-based tutoring systems on educational experiences. By leveraging advancements in natural language processing, machine learning, and human-computer interaction, CITS have the capacity to accommodate diverse learning styles and enhance knowledge acquisition in educational settings.

The integration of conversational agents and natural language interfaces in ITS has been a key area of advancement. For example, AutoTutor~\cite{Graesser2004AutoTutorAT, Nye2014AutoTutorAF} and Watson Tutor~\cite{DHelon2019INTERACTIVELI} are tutors with dialogue in natural language, emphasizing the role of conversational interactions in enhancing learning experiences. PUMICE~\cite{Li2019PUMICEAM} is a multimodal agent that learns concepts through natural language and demonstrations, emphasizing the importance of integrating multiple modalities in intelligent tutoring systems for effective communication. Furthermore, SUGILITE~\cite{Li2020InteractiveTL} integrates GUI-grounded natural language instructions to develop a conversational assistant for Android, showcasing the fusion of interactive task learning with real-world applications. By integrating conversational breakdown repairs, CITS can provide users with a seamless learning experience by addressing communication challenges through multi-modal interfaces~\cite{Li2020MultiModalRO}. More recently, LLMs capacity to generate coherent responses, understand context, and adapt to user input has opened new avenues for educational applications\cite{AlbornozDeLuise2023OnUC, Kasneci2023ChatGPTFG}.

The effectiveness of CITS in supporting personalized learning has been a subject of interest in educational research. Zinn~\cite{Zinn2013AlgorithmicDF} discusses algorithmic debugging for intelligent tutoring, focusing on improving diagnosis and problem-solving support within ITS, underscoring the importance of multiple models for enhanced learning outcomes. Additionally, Aljameel et al.~\cite{Aljameel2018CanLC} evaluate the application of LANA CITS in supporting learning for autistic children, showcasing the potential of Conversational Intelligent Tutoring Systems to cater to diverse learning needs. It has been shown that CITS provides the capability to encourage mentees, especially when learning knowledge within the STEM domain \cite{Feng2021ASR}. Moreover, Akyuz~\cite{Akyuz2020EffectsOI} explores the impact of Intelligent Tutoring Systems on personalized learning, highlighting the positive contributions of ITS in enhancing student performance and time management. \system builds upon these prior works, providing support for authoring and assessing skill mastery to build student models for adaptive teaching and leveraging advances in LLM applications in chatbots to build conversational capabilities.
\section{User Experience }

\begin{figure*}[t!]
  \centering
  \includegraphics[width= \textwidth]{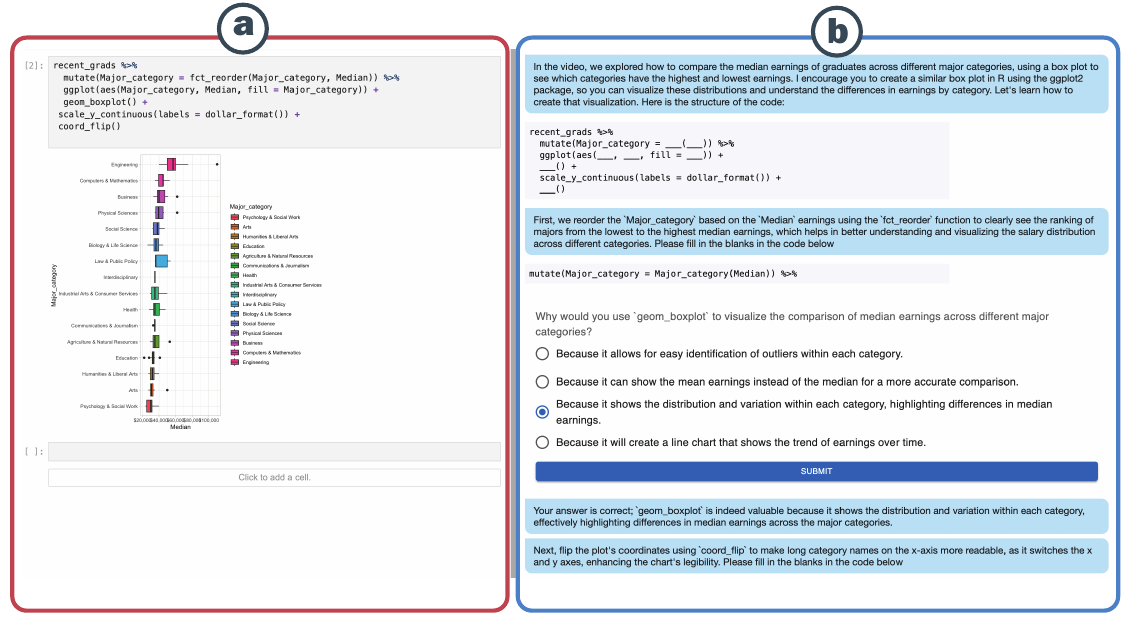}
  \caption{\system user interface as a JupyterLab Extension. (a) Code panel with cell structure, (b) \system Chat Interface with Rich Conversational Features}
  \label{fig:interface}
\end{figure*}

\system is an integrated learning environment where learners can watch programming video segments, engage in learning conversations with the AI mentor, and practice writing code. As shown in Figure~\ref{fig:interface}, \system is implemented as an extension to \textit{JupyterLab}~\cite{ProjectJupyter} --- an interactive computational environment for notebooks, code, and data. The extension primarily consists of a rich \textit{ multi-modal} conversational chat panel added to the main JupyterLab interface. The default configuration consists of a notebook panel on the left and the \system chat panel on the right. However, learners can adjust the position of the panels at will. The chat panel and the code cells in the notebook are linked and controlled by \system. To better understand how \system transforms programming videos into a one-on-one mentoring experience, let us follow Leon, a student in a Data Science undergraduate program interested in learning Visual Exploratory Data Analysis (EDA) using the `R' programming language~\cite{RProject2024}. Leon has familiarity with R programming and has installed the \system extension on his JupyterLab desktop application using a one-click installer.

\subsection{Video Context}
For this example, Leon will use the YouTube video ``Tidy Tuesday Screencast: analyzing college major \& income data in R~\footnote{\url{https://www.youtube.com/watch?v=nx5yhXAQLxw}}'' by Dave Robinson. To provide context, Dave Robinson is a well-known figure in the data science community, particularly known for his involvement with the TidyTuesday project and his screencasts on YouTube~\cite{RobinsonTidyTuesdayYouTube}. TidyTuesday~\cite{tidytuesday} is a weekly data project aimed at the R community that encourages participants to apply their data wrangling and visualization skills to a new dataset every week. While not intended as tutorials, Dave regularly publishes his live coding sessions on YouTube, which embodies his deep expertise in data science. Leon has attempted to learn by directly watching the video, but self-regulation~\cite{zimmerman2011self} -- in which the learner actively manages their own learning experiences through planning, monitoring, and evaluating their understanding -- has been challenging. Our goal is for Leon to leverage the expert knowledge in the video to improve his EDA skills using the apprenticeship model approach. 

\subsection{Apprenticeship Learning with \system}
Upon loading the chat panel, the \system chatbot prompts Leon to upload a configuration file for the topic he wishes to learn along with links to the video and code (we detail the file format and creation in Section~\ref{sec:expertinput}). Based on the configuration, learning EDA involves the following \textit{learning goals}: gaining familiarity with the dataset, data cleaning, forming hypotheses or visualization intent, visualization construction, visualization refinement, and visualization interpretation and insight generation~\cite{tukey1977exploratory}. Using the configuration \system instantiates a student model for Leon with a skill level of `novice' for all of the goals. 

To get the tutoring session started, \system identifies that gaining \textit{familiarity with the data} is the first relevant segment in the video. Since this is also the first time Leon is encountering this goal, \system takes the \textit{MODELING} move and presents the video segment in which Dave explores the dataset using think-out-loud as a chat response. For goals involving \textit{declarative knowledge} such as facts, \textbf{playing the video segment} in which the expert (Dave) is exploring the dataset is a suitable move. \system instructs Leon to watch the short video clip. After learning about the dataset characteristics from the video, Leon clicks on the ``Go On'' button, indicating that he is ready to proceed. The next step in the video is creating a visualization to explore the relationships between \textit{College Major} and \textit{Income}. Since this step involves \textit{procedural knowledge} by writing code to create a box plot, the modeling move involves \system presenting the learner with the code snippet and walking them through each of the functions and parameters. \system has access to Dave's completed code along with the transcript, which is used to generate the correct code. Note that in the process, \system also automatically adds the code to the notebook cell. At the end of the \textbf{code walkthrough}, \system chatbot instructs Leon to execute the code and generate the visualization. Further, for visualization interpretation, this move again plays the relevant video segment. 

As the video progresses, Leon encounters a second instance of a bar chart visualization. In this case, \system takes the \textit{COACHING} move and provides Leon with a \textbf{code template} for the bar chart with key functions and parameters left blank. Then, based on the visualization intent, \system guides Leon in filling out the blanks through conversation and interactivity. For instance, the blanks in the code template are clickable, and Leon can select from the dropdown list of options. \system provides feedback in case of incorrect selection. Once completed, \system adds the code to the notebook, which can be executed. For visualization interpretation, the coaching involves specific guidance to Leon on what to attend to in the visualization. \textit{SCAFFOLDING} and \textit{ARTICULATION} moves for declarative knowledge involves \textbf{multiple choice questions}, asking Leon to pick the right answer, and providing feedback. These moves for procedural knowledge could involve \system asking Leon to directly write the code or interactively explore the dataset to formulate hypotheses and provide \textbf{feedback}. As Leon is writing code in the notebook, \system evaluates the code for errors and assists Leon in \textbf{fixing code errors}. Alternately, in \textit{REFLECTION} move, \system can ask the student to assess the issues on their own. Lastly, \textit{EXPLORATION} allows Leon to go beyond what is contained in the source video, such as generating and testing hypotheses that are not covered in the video. Throughout these moves, the \system constantly updates the underlying student model to capture the content and learning goals Leon has encountered. 

In summary, \system takes the source video and applies different mentoring moves from the \framework framework and interactively walks Leon through the video. In the learning conversations with \system, Leon takes a learning-by-doing approach to develop his EDA and visualization skills and receives active feedback across different EDA tasks. In the process, Leon can also ask clarifying questions and work at his own pace to regulate his learning.

\section{System Architecture}

\begin{figure*}[t!]
  \centering
  \includegraphics[width= \textwidth]{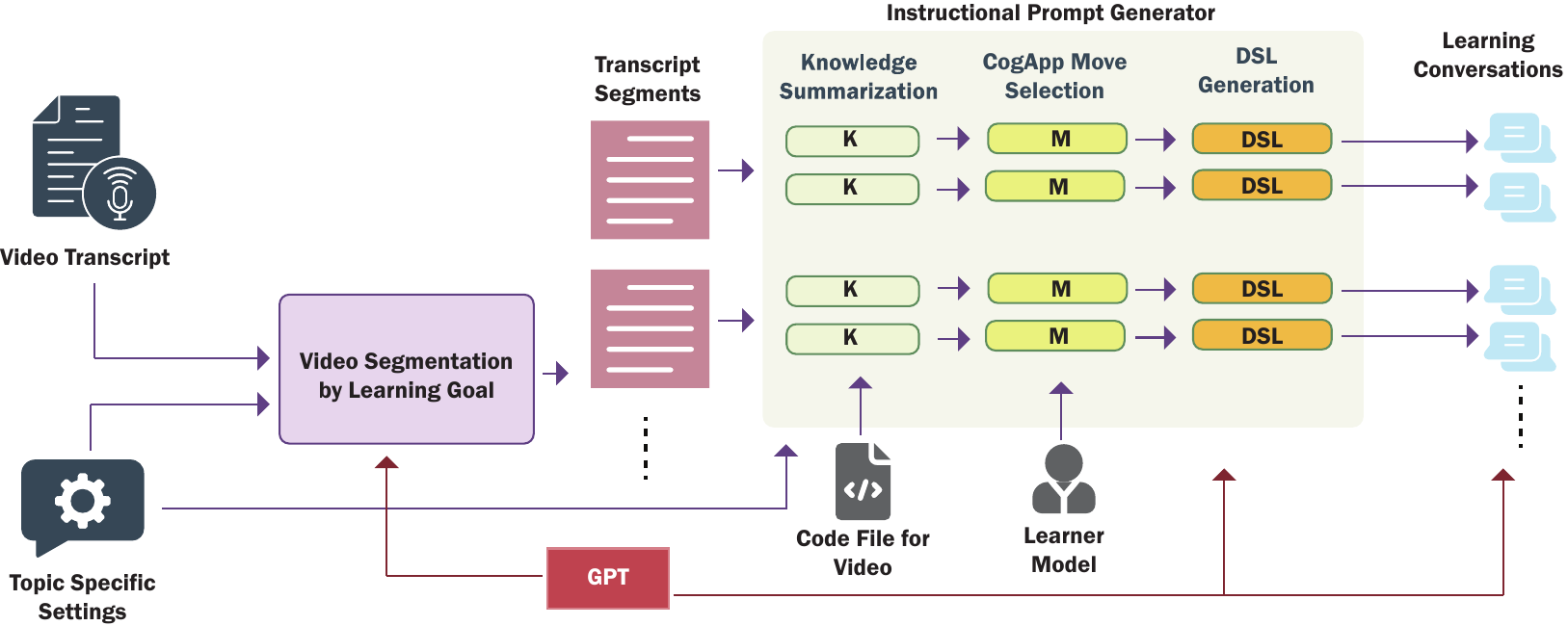}
  \caption{Overview of \systems System Architecture. From Left to Right: The video Transcript is segmented based on learning goals, and then for each segment, the knowledge is summarized based on knowledge type. Next the appropriate \framework mentor move is selected for the knowledge and using the student model. Finally a DSL representation is produced to prompt GPT for generating the conversation. }
  \label{fig:arch}
\end{figure*}

\system implements a set of LLM prompting techniques that yield conversational utterances grounded in the \framework framework. As shown in Figure~\ref{fig:arch}, our system consists of (1) a video segmentation module that \textit{slices} long videos based on learning goals using a prompt chaining approach, (2) a DSL generator that produces the \textit{context} for the LLM to generate utterances in interaction with the learner, and (3) a student model to capture the student's learning progress and inform appropriate mentor \textit{moves}. Here, we provide technical details about each of these modules along with challenges and prompt engineering strategies to make the generated conversations \textbf{consistent} and \textbf{controllable}:

\subsection{Video Segmentation by Learning Goals}

To implement the \framework framework, we need to be able to distinguish specific learning goals in the video and map them to mentoring strategies. While GPT-4 has strong summarization capabilities~\cite{zhang2024benchmarking}, at the time of this writing, GPT-4 could not yet directly perform long-context semantic classification. First, when segmenting the transcript by just providing the learning goals, we observed that even small references to other learning goals within a main segment (e.g., mention of data or hypothesis in segmentation about visualization construction) led GPT to distinguish it as having two learning goals, thus affecting the classification results. Additionally, current transcripts are segmented and timestamped based on pauses in speech or changes in speakers. While advanced ASR attempts to improve the readability of transcripts with complete sentences and the use of punctuation, it can be challenging and is not always accurate. These challenges in transcription impact the inductive capabilities of large language models.

To address these challenges, we start by defining each of the learning goals as a one-shot prompt using a single descriptive example as context. For instance, if the goal is \textit{chart interpretation}, we provide a short description that it means instances in which ``the expert interprets the visualizations and discusses the implications of the visualization, drawing conclusions, and theorizing about the underlying trends or patterns in data.'' In our iterative trials, GPT is able to reasonably generalize and infer goals based on these descriptions. In addition, we provide few-shot examples of the segmented videos to produce distinct summaries of multiple occurrences of the learning goals in the video. Given this context, we utilize the chain-of-thought (CoT) prompting technique ~\cite{wei2023chainofthought} to define the \textit{Video Segmentation by Learning Goal} (VSLG) algorithm defined as follows: 

Let \( V = \{v_1, v_2, ..., v_T\} \) be the set of all sentences in the video transcript, where \( T \) is the total number of sentences, and \( L = \{l_1, l_2, ..., l_M\} \) be the set of learning goals. The transcript is segmented into \( S = \{s_1, s_2, ..., s_N\} \), where \( s_n \) corresponds to one segment for one of the learning goals \( l_m \). \( N > M \), so that each learning goal has multiple segments. We define three functions executed sequentially in Algorithm~\ref{alg:VSLG}, which mimics the cognitive steps an individual might take when classifying video clips semantically and provides a structured methodology for LLMs to segment video content by learning goals.

\begin{algorithm}
\caption{VSLG - Video Segmentation by Learning Goal}\label{alg:VSLG}
\begin{algorithmic}[1]
    \Require Video transcript sentences \( \mathcal{V} \), Learning goals \( \mathcal{L} \)
    \Ensure Video segments by learning goals \( \mathcal{S} \)
    \State \( \mathcal{S'} = \{s'_1, s'_2, ..., s'_N\} \gets \textbf{Summarize}(\mathcal{V}, \mathcal{L}) \)
    
    \Comment{summarizes transcript of each learning goal, yielding a set of summary points of all segments without timestamps.}
    \State \( \mathcal{R} \gets \textbf{Retrieve}(\mathcal{S'}, \mathcal{V}) \)
    
    \Comment{retrieves sentences from the video transcript for each summary point \( s'_n \).}
    \State \( \mathcal{S} \gets \textbf{Rearrange}(\mathcal{R}) \) 
    
    \Comment{determines the range of segments based on the timestamps of sentences and sorts them by time.}
\end{algorithmic}
\end{algorithm}

Below is a snippet of the segmentation process using this approach ("start" and "end" are the start time and end time of this segment in seconds):

\begin{promptbox}
\textbf{Input: [video transcript]}

\noindent
\textbf{[After the first step]}
\begin{lstlisting}[language=json, breaklines=true]
[...
  {"category": "Visualize the data",
   "summary": "David decides to explore the median salaries by creating a histogram to understand the distribution of median earnings across majors.'},
...]
\end{lstlisting}
\noindent
\textbf{[After the second step]}
\begin{lstlisting}[language=json, breaklines=true]
[...
  {"category": "Visualize the data",
   "sentence": "alright so I take a look at every synchros now that it picked something I'm interested in...'},
...]
\end{lstlisting}
\noindent
\textbf{[After the final step]}
\begin{lstlisting}[language=json, breaklines=true]
[...
  {"category": "Visualize the data",
   "start": 435.23, "end": 461.93},
...]
\end{lstlisting}
\end{promptbox}

\subsection{Instructional Prompt Generator}
The next step in the pipeline involves generating prompt \textit{templates} and associated \textit{context} for each video segment. The prompts need to align with different moves in the \framework framework to drive the conversational interactions with the learner later. Applying \textit{chain-of-thought}, if an instructor were teaching, they would begin by determining the specific knowledge or content that the goal should encompass~\cite{proitz2010learning}. Then, they would strategically determine pedagogical approaches for different learner needs. Similarly, our prompting strategy is first to use the segmented video transcripts and code to summarize the knowledge to be learned, then determine the teaching method based on the knowledge type for each of the moves in \framework Framework. Finally, we map the method to predetermined actions and interaction types. By enforcing a step-by-step generation process, we emulate instructional practices and gain fine-grained control to carry out the conversation with learners progressively. Here, we detail each of these steps in generating instructional prompt templates:

\begin{table*}[ht!]
\centering
\renewcommand{\arraystretch}{1.5}
\begin{tabular}{p{2cm}|p{6cm}p{6cm}}
\hline
\textbf{Knowledge} & \textbf{Representation} & \textbf{Example} \\ \hline
\multicolumn{3}{c}{\textit{Concept Related}} \\ \hline
Declarative & [Subject] + [verb phrase] + that + \textbf{[independent clause]}. & \textit{The median income by college major shows that \textbf{majors earn a median income of over \$30K right out of college}.} \\ \hline
Procedural & To achieve/understand + [specific goal/outcome] + one must + \textbf{[actions/processes]} + [additional details] + considering/using + [relevant factors/tools]. & \textit{To understand the distribution of earnings by college major, one must \textbf{examine the histogram and identify overall trend or extreme values}, considering whether high earnings are due to the field's financial reward or influenced by factors such as low sample size and high variation.} \\ \hline
\multicolumn{3}{c}{\textit{Programming Related}} \\ \hline
Declarative & The task is + [final goal] + using + [general method/tool] + and + [additional method/technique for enhancement]. & \textit{The task is comparing the distribution of median earnings across different major categories using a box plot and adjusting the visualization for better readability and interpretation.} \\ \hline
Procedural & To achieve + \textbf{[specific goal]} + one must + \textbf{[action/verb] + [specific tool/method]} + on + [object/target] + because + [reason/purpose]. & \textit{To \textbf{achieve an ordered factor level} based on the `Median`, one must \textbf{use `fct\_reorder`} on `Major\_category`, making it easier to compare distributions.} \\ \hline
\end{tabular}
\caption{Representations of the two types of knowledge in \system. Each type of knowledge has concept-related and programming-related representations. The bold parts are later used in the student model.}
\label{tab:knowledge}
\end{table*}

\subsubsection{Summarize the knowledge in each video segment.} Grounded in the widely used revised Bloom's Taxonomy~\cite{anderson2001taxonomy, ryle1945knowing}, we focus on \textit{declarative} Knowledge (i.e., propositional facts, information, and descriptions ) and \textit{procedural} knowledge (i.e., knowledge about processes and sequences of actions) as they are most common in programming education. Through trial and error with different levels of knowledge abstractions, being too precise leads to overlapping information in the summaries, making downstream tasks more complex and less generalizable. Further, we find that in programming videos, declarative and procedural knowledge is spread across domain \textit{concepts} (e.g., objects in object-oriented programming, dataset attributes) as well as \textit{programming} constructs (visualization, sorting). Using this categorization, we use few-shot prompts to generate procedural and declarative knowledge summaries in each of the video segments. Additionally, as shown in Table~\ref{tab:knowledge}, for each knowledge-content category, we define specific summary formats. First, this ensures that we are able to map the knowledge to build and update the student model more accurately (Section~\ref{sec:studentmodel}). Second, the summary format contains the necessary information for GPT to generate conversational messages without making the knowledge content too long or too short. Third, the format allows us to link the transcript to the code. If the knowledge summary is vague, for instance, \textit{``Filter the dataset to select the top 20 rows to focus on the highest-earning majors''}, it becomes difficult to map to functions and attributes in the code for downstream interactions. We designed the summary formats iteratively through design and testing. 

\subsubsection{Select Pedagogical Strategy for each piece of knowledge.} Given the knowledge summary, the next step is for the LLM to determine the instructions grounded in the \framework Framework. If LLMs have a high degree of freedom, i.e., zero-shot prompting by giving it the framework and letting it plan the instructions, it is challenging to maintain logical consistency and coherence~\cite{liu2024trustworthy}. In our trials, we observed that with this approach, the LLM might 
use the \textit{REFLECTION} move even in the early stages and regardless of the knowledge to be learned. On the other hand, rigidly mapping specific teaching methods to particular knowledge can constrain \systems ability to adapt to unforeseen content and interactions, and the monotony may lead to disengagement. Our goal is to support a dynamic instructional approach within the boundaries of \framework frameworks. 

To do this, our prompt engineering follows the following three principles in the \framework framework to select an appropriate mentor move: \textbf{ (1) Global before local skills.} The recommendation is to have learners build a conceptual map before attending to the details of the terrain. To comply with this principle, \system chooses \textit{Modeling} or \textit{Coaching} as the first move to describe the tasks explored in the video segment or demonstrate the end-to-end process of solving the tasks. This move would then be followed by other methods to guide students in using more advanced problem-solving strategies. \textbf{ (2) Increasing complexity.} With this principle, the difficulty of the task gradually increases, so each new task introduces additional layers of skills and concepts essential for mastery. We implement this through a combination of prompt engineering and student modeling, described later. \textbf{(3) Increasing Diversity.} This means providing students with a broad range of problems and contexts, which compels them to apply their skills and strategies in different ways \cite {bransford1999chapter}. To optimize the learning experience, it is essential to diversify teaching methods, even when students are engaging with different video segments that share the same learning goals. For example, if a student learned how to make a histogram by answering a multiple-choice question in the past video segment, next time, the \system will ask them to fill in the code blanks to make a box plot. Varying the instructional approach could maintain student interest and cater to different learning styles.

The process is detailed in Algorithm~\ref{alg:knowledge_teaching}. Here is one example in the prompt (additional details in the Appendix):

\begin{algorithm}
\caption{Knowledge Summary and Pedagogical Move}\label{alg:knowledge_teaching}
\begin{algorithmic}[1]
    \Require Video segments \( S \), Cognitive Apprenticeship framework \( \mathcal{F} \), code blocks \( C \), student model for each knowledge \( p_{k} \)
    \Ensure Summarized knowledge \( K \) and teaching methods \( M \)

    \State Initialize knowledge list \( K \gets \emptyset \)
    \State Initialize teaching method list \( M \gets \emptyset \)
    
    \Function{SummarizeKnowledge}{$S$}
        \For{each segment \( s \in S \)}
            \State \( K_s \gets \) Generate knowledge snippets for \( s \) with \( C_s \)
            \State Sort \( K_s \) in order of appearance within \( s \)
            \State \( K \gets K \cup K_s \)
        \EndFor
        \State \Return \( K \)
    \EndFunction
    
    \Function{PlanTeachingMethods}{$K, \mathcal{C}$}
        \For{each knowledge \( k \in K \)} \Comment{Apply the principles}
            \State Determine the index, complexity, diversity of \( k \)
            \State \( M_k \gets \emptyset \) \Comment{Initialize the set of methods for knowledge \( k \)}
            \If{\( index <= 1 \)}
                \State \( M_k \).add(\textit{Modeling}/\textit{Scaffolding}) 
                
                \Comment{Start with global skills and task overviews}
            \ElsIf{\( p_{k} <= 0.5 \)}
                \State \( M_k \).add(\textit{Scaffolding}/\textit{Coaching}/\textit{Articulation})
                
                \Comment{Depends on which one has not been used}
                \State \( M_k \).add(\textit{Reflection})
                \Comment{Encourage self-assessment}
            \ElsIf{\( p_{k} > 0.5 \)}
                \State \( M_k \).remove(\textit{Scaffolding})
            \EndIf
            \State \( M \gets M \cup M_k \)
        \EndFor
        \State \Return \( M \)
    \EndFunction

    \State \( K \gets \Call{SummarizeKnowledge}{S} \)
    \State \( M \gets \Call{PlanTeachingMethods}{K, \mathcal{C}} \)
\end{algorithmic}
\end{algorithm}

\begin{promptbox}    
\textbf{Input: [Knowledge of a concept-related segment]}

\noindent
\textbf{Output:} 
\begin{lstlisting}[language=json, breaklines=true]
{"knowledge": "To interpret the differences in median income by college major...",
 # Start with Modeling or Scaffolding
 "actions": [{"method": "Scaffolding"}]},
{"knowledge": "The chart on median income by college major contains...",
 # Increase diversity by using more methods
 "actions": [{"method": "Articulation"}, 
             {"method": "Reflection"}]},
{"knowledge": "To draw final conclusions from the chart...",
 # Increase complexity followed by Reflection
 "actions": [{"method": "Coaching"}, 
             {"method": "Reflection"}]}
\end{lstlisting}
\end{promptbox}

\subsubsection{Generating a DSL for multi-turn conversation.} Given a knowledge summary and pedagogical move, we still need a reliable way to operationalize it in generating \system chatbot conversations. Zero-shot prompting will not work. First, teaching moves are somewhat abstract principles and, without specificity and context, can lead LLMs to produce lengthy and unfocused text, hindering learning effectiveness. Second, the conversation should be generated more actively and in a step-by-step manner. Typically, chatbots respond reactively to user inputs rather than initiating a dialogue that steers the learning process. Further, the timing of messages is also critical: some should be delivered promptly to maintain the learning flow, while others should be contingent upon the learner's actions to ensure relevance and engagement.

To support these learning conversation needs, we developed a domain-specific language (DSL) that encompasses all the necessary information for directing chatbot interactions within a learning scenario. As a prerequisite, for each pedagogical move, we need experts to provide at least one action type and response interaction for programming-related and concept-related knowledge (see Section~\ref{sec:expertinput} for details). Our DSL incorporates information from earlier processing steps and the action information in the pipeline to orchestrate the conversation. As shown in Algorithm~\ref{alg:DSL}, we first map the expert-provided \textit{actions} and \textit{interactions} for each pedagogical move. Then, we used few-shot prompting to generate a \textit{prompt} and \textit{need-response} based on each pair of action-interaction, which is used to drive the LLM to generate an utterance and detect whether it needs the user to respond to generate the next utterance. The \textit{parameters} is a list of parameters required in the ``prompt'', such as knowledge, code block, student's answer, etc. The sequence of messages depends on the knowledge and the actions that help students learn the knowledge. They are stored in a queue, and each time a message is generated, the head of the queue is removed. The full structure can be formed before uploading to the chatbot, thus avoiding calling the prompt pipeline to save time during the conversation. The algorithm and part of an example DSL structure is shown below:

\begin{algorithm}
\caption{DSL Structure for Chatbot Message Generation}\label{alg:DSL}
\begin{algorithmic}[1]
    \Require Methods Set \( \mathcal{M} \), Actions \( \mathcal{A} \), Interactions \( \mathcal{I} \), Parameters list \( \mathcal{P} \), student behavior \( \mathcal{B} \), knowledge \( K \), student model \( p_{K} \)
    \Ensure Chatbot interaction
    \Function{GetDSL}{$a, i$}
        \State \textbf{Description}: Generates DSL components for \( \mathcal{M} \).
        \State \( prompt \gets \) Use \( i \) to generate the prompt for action \( a \) 

        \Comment{"Use [interaction] to + definition of the method"}
        \State \( needResponse \gets \) Determine if requires user response
        \State \( \mathcal{P} \gets \) Extract parameters required for the prompt
        \State \Return \( \{prompt, needResponse, \mathcal{P}\} \)
    \EndFunction
    \State Initialize message queue \( \mathcal{Q} \)
    \For{each \( m \in \mathcal{M} \)}
        \For{each \( (a, i) \in \mathcal{A} \times \mathcal{I} \)}
            \State \( \{prompt, needResponse, \mathcal{P}\} \gets \Call{GetDSL}{a, i} \)
            \State Enqueue \( \{m, a, i, prompt, \mathcal{P}, needResponse\} \) into \( \mathcal{Q} \)
        \EndFor
    \EndFor
    
    \While{not \( \mathcal{Q}.empty \)}
        \State \( message \gets \mathcal{Q}.dequeue \)
        \State \( \textbf{Chatbot}.send(message.prompt, message.\mathcal{P}) \)
        \If{\( \textbf{message}.needResponse \) is \textbf{false}}
            \State \textbf{continue}
        \Else
            \State \( \textbf{Chatbot}.wait(\textbf{Signal}_{response}(t)) \)
            \State \( p_{K} \gets \textbf{UpdateStudentModel}(\mathcal{B}, K, p_{K})\)
        \EndIf
    \EndWhile
\end{algorithmic}
\end{algorithm}

\begin{promptbox}
\begin{lstlisting}[language=json, breaklines=true]
"Visualize the data - 509": [
{
  "knowledge": "Use `geom_boxplot` on ...",
  "actions": [
  {
    "method": "Coaching",
    "action": "Prompt the student to use {interaction} to practice the knowledge",
    "interaction": "fill-in-blanks",
    "prompt": "[Use one sentence to prompt the student to fill in the {code-line-with-blanks} below]",
    "parameters": ["code-line-with-blanks"],
    "need-response": true
  },
  ...
  ]
},
...
]
\end{lstlisting}
\end{promptbox}

\subsection{Student Behavior Monitoring} \label{sec:studentmodel}
A novel aspect of \system is that the conversation is not only driven by the video context but also by modeling the learner. Here, we cover key details on student modeling, including the timing of behavior evaluation, updating student models, and the influence of these models on the prompt pipeline:

\subsubsection{Monitor Trigger Conditions.} Based on literature~\cite{vanlehn2006behavior,baker2007modeling} and trial and error, we identify a set of signals that updates the knowledge state of the student model. The signals are intended to ensure optimal monitoring frequency; excessive instructor intervention can distract students and slow system responsiveness, while infrequent updates to student models may delay crucial learning feedback. We implement a signal function \( \textbf{Signal}(t) \). When it is activated, \system updates the student model. We define the four monitor trigger conditions as follows:

\begin{itemize}
    \item \textit{Post-viewing Guidance}: Initiated by \( \textbf{Signal}_{video}(t) \), where the instructor provides guidance based on the video.
    \item \textit{Progression Cue}: Triggered upon completion of an instruction \( \textbf{Signal}_{response}(t) \), signaling the next step of guidance.
    \item \textit{Corrective Feedback}: Activated by \( \textbf{Signal}_{error}(t) \), when the code execution fails, prompting improvement suggestions.
    \item \textit{Responsive Assistance}: In response to \( \textbf{Signal}_{help}(t) \), the instructor answers questions or provides additional help.
\end{itemize}

\subsubsection{Student Model Update.} Although many ITS use question-answer pairs to test student learning outcomes~\cite{Das2021AutomaticQG}, evaluating students' performance on specific problems cannot directly represent students’ knowledge models~\cite{Shafiyeva2021AssessingSM}. Utilizing a part of the \textbf{Knowledge} as outlined in Table \ref{tab:knowledge}, we construct a statistical model to emulate the student's mental state using Bayesian Knowledge Tracing (BKT) based on a Hidden Markov Model. Each knowledge component is characterized by four parameters: \( p_{mastery}^{(t)} \), \( p_{transit} \), \( p_{slip} \), and \( p_{guess} \). These parameters are defined as follows:

\begin{itemize}
    \item \( p_{mastery}^{(t)} \): The probability that the student has mastered the knowledge at time \( t \).
    \item \( p_{transit} \): The probability of the student transitioning to a state of knowledge after an opportunity to apply it.
    \item \( p_{slip} \): The probability the student makes an error when applying known knowledge.
    \item \( p_{guess} \): The probability the student correctly applies unknown knowledge.
\end{itemize}

Our system updates the student model using Equation~\ref{eq:BKT}, based on the observed student behavior \( \text{obs} \) at time \( t \) when \( \textbf{Signal}(t) = True\)

\begin{equation}
\label{eq:BKT}
    p_{mastery}^{(t|obs)} = 
    \begin{cases}
    \frac{p_{mastery}^{(t)} \times (1 - p_{slip})}{p_{mastery}^{(t)} \times (1 - p_{slip}) + (1 - p_{mastery}^{(t)}) \times p_{guess}}  &\text{obs = correct} \\
    \frac{p_{mastery}^{(t)} \times p_{slip}}{p_{mastery}^{(t)} \times p_{slip} + (1 - p_{mastery}^{(t)}) \times (1 - p_{guess})}  &\text{obs = incorrect}
    \end{cases}
\end{equation}
\\
\system uses this formula to update the student model throughout the learning process. Every time the student is practicing a piece of knowledge, it is recorded and initialized using the BKT parameters. If the student is practicing knowledge that has been practiced with similar knowledge before, the BKT parameters of the existing knowledge will be updated. We use semantic similarity to evaluate whether similar knowledge exists~\cite{youdao_bcembedding_2023}. During the practicing process, if students respond correctly, such as choosing a correct choice in the multiple-choice question or writing the correct code line, the observation is \textit{Correct} and \system accordingly updates the parameters of that correctly practiced knowledge. On the contrary, if students respond incorrectly, \system uses the formula for incorrect to update the parameters of the incorrectly practiced knowledge. Each time a student completes a learning session, BKT parameters are saved to the database. When students watch a new video, \system will load previously saved parameters for the new learning course. The whole process is shown in Algorithm~\ref{alg:student_model}.

\begin{algorithm}
\caption{Update Student Model}\label{alg:student_model}
\begin{algorithmic}[1]
    \Require student behavior \( \mathcal{B} \), knowledge \( K \), student model \( p_{K} \)
    \Ensure the Student model gets updated
    \Function{UpdateStudentModel}{$\mathcal{B}, K, p_{K}$}
        \State \textbf{Description}: Updates the student model.
        \State Wait for \( \textbf{Signal}(t) \) before proceeding
        \For{each knowledge \( k \in K \)}
            \If{\( k \) is semantically similar to an element in \( K \)}
                \State \( p_{k} \gets \) the parameter for \( k \) in \( p_{K} \)
                \If{\( \mathcal{B} \) is correct}
                    \State Update \( p_{k} \) based on correct behavior
                \Else
                    \State Update \( p_{k} \) based on incorrect behavior
                \EndIf
            \Else
                \State Create new parameter \( p_{k} \) for \( k \)
                \If{\( \mathcal{B} \) is correct}
                    \State Initialize \( p_{k} \) based on correct behavior
                \Else
                    \State Initialize \( p_{k} \) based on incorrect behavior
                \EndIf
            \EndIf
        \EndFor
        \State \( p_{K} \gets p_{K} \cup p_{k} \)
        \State \Return updated \( p_{K} \)
    \EndFunction
    \State \( p_{K} \gets \Call{UpdateStudentModel}{\mathcal{B}, K, p_{K}} \)
\end{algorithmic}
\end{algorithm}

\subsubsection{Integration into the Prompt pipeline.} The student model is a direct reflection of the student's current knowledge and understanding. Based on prior work~\cite{Dai2019ResearchOM}, our pipeline integrates the student model for the following three tasks: 

\begin{itemize}
    \item Selecting learning goals for students showing weak mastery, indicated by \( p_{mastery}^{(t)} < 0.3 \).
    \item Reducing the recurrence of knowledge components already mastered, denoted by \( p_{mastery}^{(t)} > 0.7 \).
    \item Adapting teaching methods according to the student's knowledge mastery; for instance, favoring \textit{Coaching} over \textit{Modeling} or \textit{Scaffolding} if \( p_{mastery}^{(t)} > 0.5 \) for knowledge ``plotting a histogram to see the distribution''.
\end{itemize}

\subsection{Implementation Details}
\system is implemented as a server-client architecture. The front end is a JupyterLab plugin~\cite{ProjectJupyter} built using \texttt{TypeScript} and \texttt{React}~\cite{ReactJS}.  We make use of the provided APIs to integrate the plugin with the main computational notebook including \texttt{@jupyterlab/notebook}, \texttt{@jupyterlab/cells}, and \texttt{@lumino/signaling}. The chat UI design uses the Chat UI Kit from \textit{chatscope}~\cite{ChatScopeUIKit}, an open-source UI toolkit for developing web chat applications based on \texttt{React}. In addition to basic incoming-outgoing message functions, it also provides timestamps, ``is typing,'' and other functions in the interface. The backend is built using \texttt{Python} and uses OpenAI's API for all language model functions, specifically the \texttt{GPT-4} and \texttt{GPT-4-32k} model~\cite{openai2024gpt4}, which are OpenAI's most capable language model at the time of writing. We use a temperature of 0.2 for mentioned prompts. We use the conversation buffer memory implemented by \texttt{LangChain}~\footnote{\url{https://python.langchain.com/docs/modules/memory/types/buffer}} to store messages during the chat. The video transcript and other necessary information are fetched using YouTube's API. The original \texttt{R Markdown} code files and dataset files of videos are fetched using GitHub's API. We use \texttt{SQLite} for the database design. We use the \textit{Bilingual and Crosslingual Embedding for RAG} model~\cite{youdao_bcembedding_2023} for text classification and semantic similarity. All server functions are implemented to ensure compatibility with the JupyterLab extension.

\subsection{Expert Inputs} \label{sec:expertinput}
\begin{figure}[htb]
  \centering
  \includegraphics[width= \columnwidth]{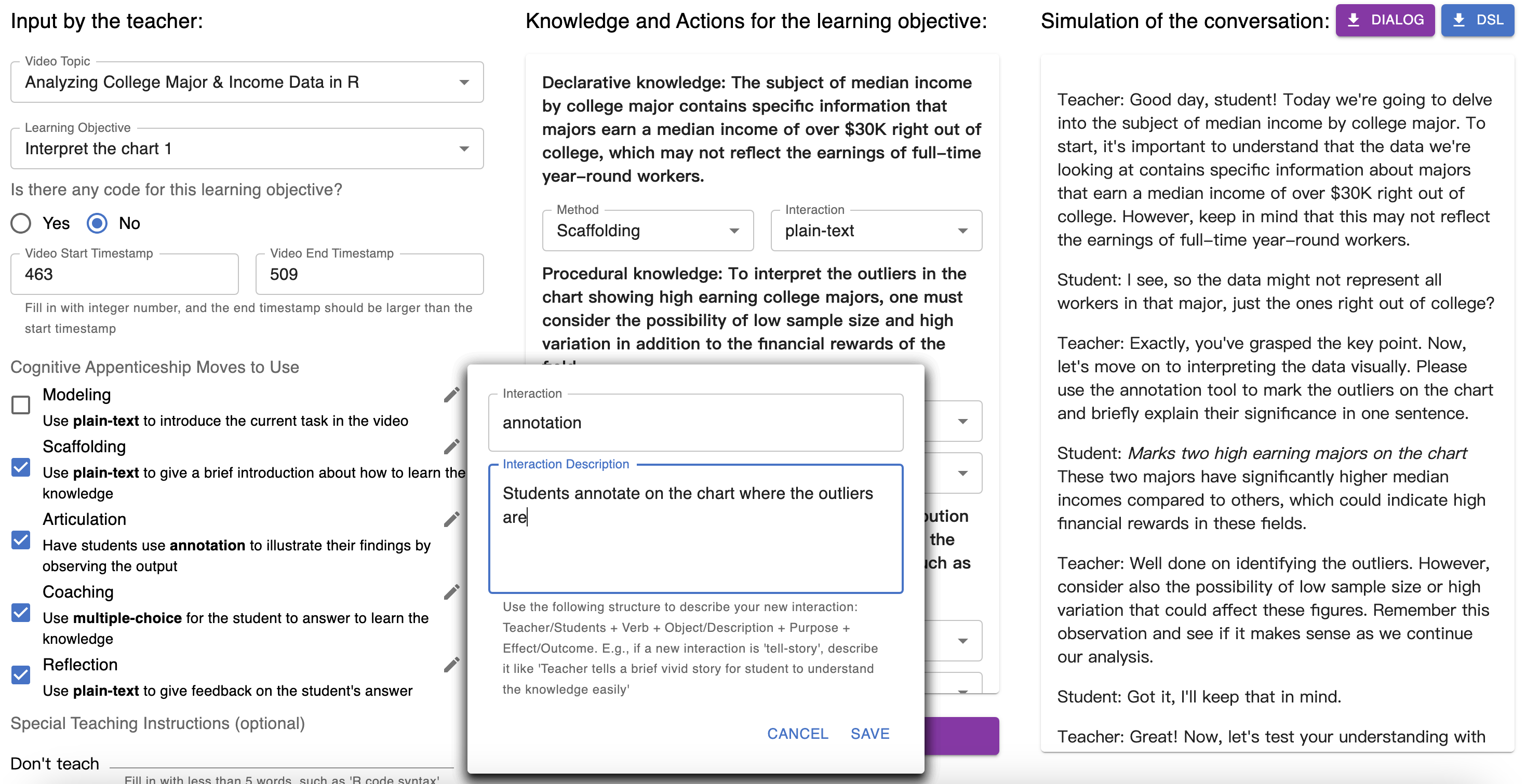}
  \caption{\system Prompt Configuration and Testing Utility}
  \label{fig:config}
\end{figure}

As mentioned above, \system relies on expert inputs to define learning goals, actions, and interactions for each of the pedagogical moves. Since these are dependent on the topic content, we abstract them out from the base implementation. This allows for the generalizability of our approach. We developed a simple web-based utility (Figure~\ref{fig:config}) to allow experts to input and configure these aspects of \system. In our current implementation, interactions are specified using well-known interventions such as  ``multiple-choice'' and ``fill-in-blanks.'' Future work can look at developing more specialized interventions. Additionally, in the utility interface, the expert can turn on or off different learning goals and conversational behavior. For instance, in a classroom setting, the instructor can configure different learning goals for students based on their needs. Based on the user experience walkthrough, if Leon excels at chart interpretation but needs to focus on creating a visualization, the instruction can turn off the interpretation goals. During active learning, \system will not generate conversations for chart interpretation. The configuration will be used in the pipeline to generate the DSL. 

\section{Technical Evaluation}
We validate our approach for three different topics (EDA, Machine Learning, and Game Development) to measure both segmentation accuracy and the generated messages for different moves in the \framework framework.  To do this, we set up a prompting pipeline by extending the utility tool to simulate both the \system chatbot and learner. Then, we ask experts to score the conversations for accuracy along the following two dimensions: 
\begin{enumerate}
    \item \textbf{Segmentation Accuracy:} Prompting must correctly categorize the video transcript into segments that correspond to the learning goals.
    \item \textbf{Controllability:} Users can adjust the knowledge, methods, actions, and interactions in the prompting pipeline. Conversation must correctly follow the user's settings to demonstrate the controllability of the pipeline.
\end{enumerate}

\subsection{Procedure}

\begin{table*}[ht]
\caption{Prompt Pipeline Labels}
\label{table:intents label}
\centering
\begin{tabular}{lp{10cm}}
\toprule
Knowledge & Example \\
\midrule
Procedural & To interpret the chart, one must analyze the histogram to identify outliers. \\
Declarative & Majors earn a median income of over \$30k right out of college. \\
\toprule
Methods & Example \\
\midrule
Modeling & I encourage you to follow along with me to create a meaningful visualization.\\
Coaching & Now, please fill in the blanks to add the box plot layer to our ggplot. \\
Scaffolding & You'll need to sort the data by median earnings in descending order. \\
Articulation & Can you tell me why it's important to consider these factors? \\
Reflection & Please compare your code with the standard code block. \\
\toprule
Interactions & Example \\
\midrule
plain-text & I'd like you to tell me what you observe about the high-earning majors. \\
multiple-choice & What could be the potential reason behind the pattern? Here are the options. \\
fill-in-blanks & Now, can you fill in the blanks to reorder the \texttt{Major\_category} based on \texttt{Median}? \\
show-code & Please execute the cell to see the visualization. \\
annotation & Please use annotation to mark the specific areas on the chart. \\
\toprule
Intent & Example \\
\midrule
Task Control & We're going to focus on interpreting a chart related to median income by major. \\
Comprehension & Visualizing the distribution of median salaries helps us to see the spread of earnings across different majors, which can be very insightful. \\
Code/Run Code & The correct line should be \texttt{recent\_grads \%>\% ggplot(aes(Median))}. \\
Feedback & You've done well in understanding the importance of each element in the plot! \\
\bottomrule
\end{tabular}
\end{table*}

The lead author, leveraging their expertise from reviewing numerous exploratory data analysis videos and familiarity with learning goals formulation, serves as the annotator for segmenting videos according to learning goals. Three additional experts recruited through our connections, proficient in each of the three video topics, served as evaluators to measure the accuracy of the intent.

\subsubsection{Evaluating Segmentation.} The lead author watched three 20-minute videos and manually segmented them using the same definitions of learning goals given to GPT. Then, we compared each segment's timestamps between the labeled data and generated data. Considering that the transcript given to GPT is discrete and the information received by watching the video is continuous, we analyzed accuracy such that the timestamp differences are within five seconds error.

\subsubsection{Evaluating Controllability.} We created an initial set of dialog intent labels (Table \ref{table:intents label}) that have a hierarchical classification scheme. This scheme also represents the hierarchical prompt pipeline. The naming rules of labels are based on the previous Pair Programming Labels~\cite{10.1145/3540250.3549127}. The ground truth data are extracted from the DSL files, and their expression is consistent with the (Table~\ref{table:intents label}). The hierarchy begins with the root \texttt{Knowledge} classifier that categorizes knowledge into procedural and declarative knowledge. The second layer is teaching \texttt{Methods} defined in the cognitive apprenticeship theory~\cite{collins1991cognitive}. We excluded \textit{EXPLORATION} in the experiment because it would add open-ended dialogues that users could not control. In the third layer, we included all the interactions used to simulate conversations. Finally, we have four types of actions that represent the four most common behaviors in our programming teaching process \cite{10.1145/3540250.3549127}.

Three expert evaluators label the data independently and separately for three topics of video (each topic has two videos): exploratory data analysis (124 utterances), machine learning (89 utterances), and game development (64 utterances). Each message is annotated with four labels. For example, the evaluator would annotate this message: ``Now, let's delve deeper. I have a multiple-choice question for you: What could be the potential reason behind the pattern of high earnings for certain majors? A) The field's inherent financial reward, B) Low sample size, C) High variation in the data, or D) All of the above.'' with the following label: ``Declarative - Articulation - multiple-choice - Comprehension''. Using this method, the evaluators annotated a total of 277 dialogue messages from the instructor. Finally, we compared the labeled data with the ground truth data with the precision, recall, and F1-score metrics to evaluate the system's accuracy, providing an understanding of how well the system's output aligns with the expected results.

\subsection{Results}

\begin{table*}[ht!]
\centering
\caption{Comparison of the intent performance across three video topics: exploratory data analysis (EDA), machine learning (ML), and game development (Game) using the precision, recall, and F1-score metrics.}
\begin{tabular}{|l|ccc|ccc|ccc|ccc|}
\hline
\multirow{2}{*}{Topic} & \multicolumn{3}{c|}{Knowledge} & \multicolumn{3}{c|}{Method} & \multicolumn{3}{c|}{Action} & \multicolumn{3}{c|}{Interaction} \\ \cline{2-13} 
                       & Prec.   & Recall  & F1   & Prec.  & Recall & F1   & Prec.   & Recall  & F1   & Prec.    & Recall   & F1    \\ \hline
Total                  & 0.791       & 0.787   & 0.789        & 0.814      & 0.809  & 0.807        & 0.902       & 0.895   & 0.896        & 0.970        & 0.968    & 0.968         \\
EDA                    & 0.810       & 0.806   & 0.808        & 0.849      & 0.815  & 0.818        & 0.896       & 0.871   & 0.872        & 0.984        & 0.984    & 0.984         \\
ML                     & 0.795       & 0.798   & 0.796        & 0.825      & 0.809  & 0.807        & 0.900       & 0.899   & 0.899        & 0.956        & 0.944    & 0.947         \\
Game                   & 0.827       & 0.781   & 0.792        & 0.808      & 0.813  & 0.798        & 0.954       & 0.953   & 0.953        & 0.973        & 0.969    & 0.969         \\ \hline
\end{tabular}
\label{tab:performance_metrics}
\end{table*}

\subsubsection{\system has acceptable segmentation results.} Video transcript segmentation by learning goals algorithm achieved an overall accuracy of 73.7\% within a five-second margin, which is an acceptable performance for educational content. However, the observed trend of decreasing accuracy with longer video duration suggests that the algorithm's effectiveness in identifying precise segmentation points diminishes over time. This observation may be attributed to the increasing complexity and variability of the content as the video progresses, making it challenging to maintain a consistent level of summarizing the video content and retrieving the relevant sentences. To alleviate this problem and improve segmentation accuracy, it could be helpful to split the video into smaller clips (e.g., every 10 to 12 minutes) and then segment by learning goals. This approach will limit the amount of content the algorithm needs to process at once, potentially reducing context load and allowing for more accurate identification of learning goals and corresponding segmentation points.

\subsubsection{\system is adept at interpreting user intent.} As shown in Table~ \ref{tab:performance_metrics}, the \textit{Total} row provides an aggregate view of \system's performance across all video topics while the other rows contain metrics of each video topic. The Knowledge column reflects \systems capacity to extract procedural and declarative knowledge from the videos. This is a critical first step as it sets the foundation for subsequent instructional design. Precision and recall scores for different video topics are around 0.791 and 0.787, indicating that \system is relatively accurate in generating messages according to the given knowledge, without containing misleading information (high precision), and without missing important content (high recall). The slightly lower performance (recall equals 0.781) in game development videos may suggest the presence of more diverse or ambiguous content in such videos, which poses a greater challenge for knowledge extraction.

The \textit{Method} column represents \system's ability to generate information that is well aligned with a defined pedagogical move. Precision and recall are also consistent for this category (difference between highest Recall to lowest is 0.006), with EDA videos showing better performance. This higher score suggests that \system is more effective at following the given methods to generate appropriate messages in an EDA environment, which may be due to the inherently interactive and action-oriented characteristics of EDA tasks. The \textit{Action} column metrics reflect how \system translates interactions into specific messages or commands. This step also shows high accuracy (over 0.900), indicating that \system can reliably trigger intended actions based on user's setting. The higher scores in game development (0.954) suggest \system's action messaging is particularly effective at identifying factual content and responding to action-driven commands in game scenarios, which may involve a mix of dynamic and static content requiring precise system responses. 

In the \textit{Interaction} column, the high scores (all around 0.950) signify \system's proficiency in generating messages based on the selected interactions. This step is crucial as it determines how students will engage with the material and how users can personalize interactions based on their needs. The near-perfect performance on all three topics reflects \system's robustness in seamlessly converting specified interactions into actual conversational utterances.

Along the process of Knowledge-Method-Action-Interaction, the scores show an upward trend (e.g., precision is 0.791-0.814-0.902-0.970 respectively). The ascending scores across these stages can be understood as a natural outcome of the instructional guidance generation process, which moves from the broad and general to the specific and interactive. At each step, \system leverages the work done in the previous phase, refining and sharpening its outputs. 

\section{User Study}
We conducted a within-subject study to evaluate the effectiveness of \system's active guidance and feedback. Concretely, we gathered user feedback on \system's support for (1) knowledge improvement in several learning goals pertaining to EDA and (2) the overall usability and cognitive load while using \system.

\subsection{Method}
We recruited $16$ participants who were proficient in the R programming language but novices in EDA through our personal networks and social media messaging. The individual sessions were conducted via Zoom and lasted approximately 90 minutes. We compensated participants with \$50 for the time. At the start of the session, participants were given instructions to install the extension on their own system or remotely control \system deployed on the lead author's laptop. Most participants opted for the latter as they also had to install JuypterLab, pip, and R on their machines, which can be time-consuming.   

In each session, participants were asked to first fill in a pre-test questionnaire to assess their prior knowledge of EDA topics. The questionnaire had five questions, including proposing a hypothesis, visualizing the data, and interpreting the chart. For example, a question for visualizing the data is ``For the problem 'What are the countries with the highest and lowest student/teacher ratios,' which visualization technique is most appropriate?''. The questions were made by the first author based on a YouTube video ``Tidy Tuesday Screencast: analyzing student/teacher ratios and other country statistics~\footnote{\url{https://youtu.be/NoUHdrailxA?si=8fziI_chqWRWFwnv}}''. Next, participants were randomly assigned to one of two conditions: (1) watch the video directly on YouTube and use an IDE of their choice and ChatGPT to support a learning-by-doing approach, or (2) use \system environment. Each condition was assigned a different video from Dave Robinson's YouTube channel~\footnote{\url{https://youtu.be/nx5yhXAQLxw?si=g6YmEUDeL3-vZ3Cw}}~\footnote{ \url{https://www.youtube.com/live/Kd9BNI6QMmQ?si=JqXX5AbZ3r2fnLiG}}. For the \system condition, participants also received a 15-minute tutorial on the main features and could try out the system on their own before proceeding to the test tasks. In the second half of the study, participants worked on the other condition, allowing for comparative feedback between watching videos with and without \system.

Finally, participants were given a post-test quiz to fill in, in which there were five questions to test their knowledge of the same learning goals of EDA after learning. The questions were based on a new YouTube video ``Tidy Tuesday Screencast: Analyzing Horror Movies in R~\footnote{\url{https://www.youtube.com/live/Eucfn-KY-t0?si=sEHUyGtZ9E2Vr0eJ}}''. Each participant then completed a usability questionnaire \cite{doi:10.1177/154193129203601617}, a cognitive load questionnaire \cite{doi:10.3758/s13428-013-0334-1}, and left optional open-ended comments on their learning and potential improvements. The study materials are provided in the supplement. 

\subsection{Results}

\subsubsection{Practice with \system results in improvements in EDA tasks.} Because the questions of each learning goal are independent (i.e., no overlap), we conducted a paired \textit{t}-test (Table \ref{tab:goals_analysis}) for the questions in each learning goal based on the results from the pre-test quiz and post-test quiz. Using \system has certain improvements in EDA tasks, but there are varying degrees of improvements for different learning goals. In the case of ``Interpret the chart,'' the mean score increased from 20.69 to 28.44, which was statistically significant, indicating that \system was particularly effective for this learning goal (p < 0.05). This may mean the teaching methods were well suited to enhance participants' competency in interpreting graphs. For ``Visualize the data,'' the mean score increased from pre- to post-test (10 to 14.38), although this improvement did not reach statistical significance (p = 0.150). ``Propose a hypothesis'' showed a slight increase in the mean score from 31.25 to 33.75, but it did not reach statistical significance (p = 0.544). Overall results indicate a positive trend in these two learning goals that may become significant with an increased quiz size or longer learning sessions.

\begin{table}[ht]
\centering
\caption{Paired \textit{t}-test results across the learning goals}
\begin{tabular}{
  l  
  S[table-format=2.2]  
  S[table-format=-1.2]  
  S[table-format=1.3]  
}
\toprule
{Learning Goal} & {Mean Diff} & {t-statistic} & {p-value} \\
\midrule
Visualize the Data   & 4.38       & -1.52         & 0.150   \\
Interpret the Chart  & 7.75       & -2.26         & 0.039   \\
Propose a Hypothesis & 2.50       & -0.62         & 0.544   \\
\bottomrule
\end{tabular}
\label{tab:goals_analysis}
\end{table}

\subsubsection{Practice with \system shows higher engagement in watching videos.} 
\begin{figure*}[ht]
\centering
\includegraphics[width=0.95\textwidth]{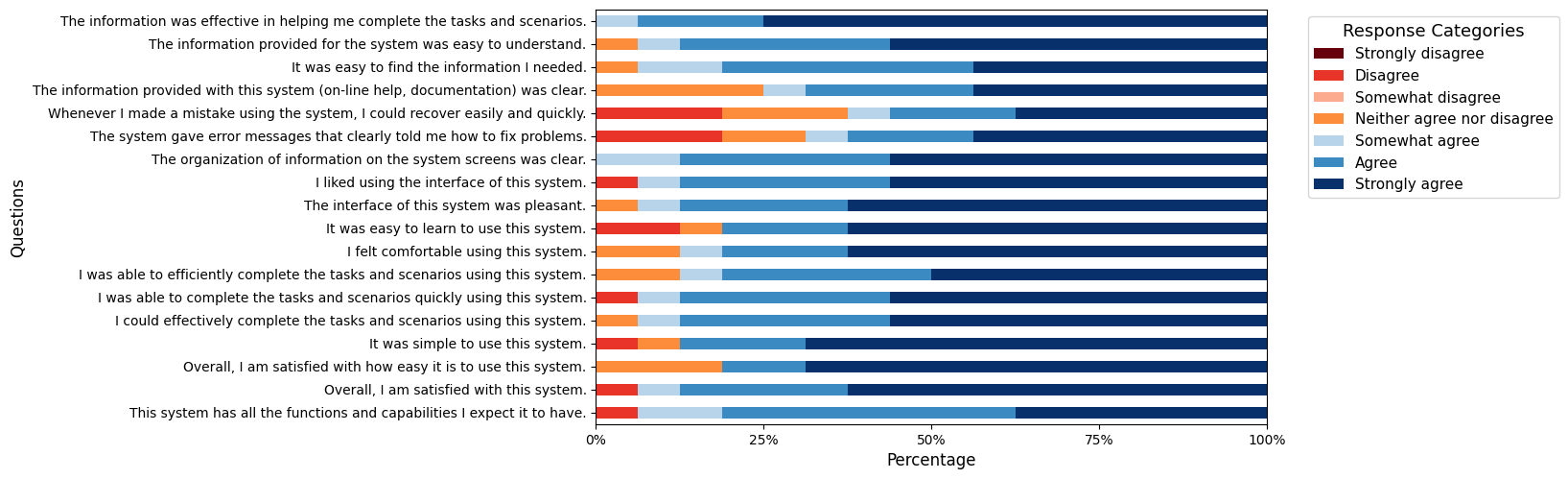}
\caption{Participant Responses to Post-Study System Usability Questionnaire.}
\label{fig:usability}
\end{figure*}

Across all sessions, participants responded positively to using \system to learn programming through watching videos. Although 75\% of the participants expressed that they typically practice programming while watching videos (i.e., learning by doing), more than half felt there was not enough explanation of the code or how to debug it (66.7\%), and there were problems applying the concepts in practice due to a lack of detailed examples (58.3\%). By contrast, 87.5\% agreed or strongly agreed that they like using \system's interface, and 81.25\% agreed or strongly agreed that it was easy to learn to use the system (Figure \ref{fig:usability}). According to P2, when watching a long instructional video, the hardest thing is to stay focused. Instead, using \system encourages them to become more immersed in watching videos and practicing: \textit{``if I want to answer the questions or fill in the code blanks correctly, I should watch the video clip carefully to avoid multiple viewings.''}(P2) Similarly, P16 commented: \textit{``I feel like I can actually practice instead of just watching videos numbly.''} 

\subsubsection{\system demonstrates a better alternative to ChatGPT.} Based on feedback from using both \system and their typical IDE alongside ChatGPT, although using ChatGPT to practice is quite common among participants (93.8\%), over half (68.8\%) point out it is hard to express their needs clearly, which led to frustration. In contrast, 93.75\% of participants believed that the information provided by \system effectively helped them complete tasks and scenarios, and 68.75\% of participants thought that the information provided by \system was clear. For example, P10 commented that \system does not generate long and tedious messages, which means it is easier to understand: \textit{``[\system] doesn’t give super long messages like ChatGPT, but it also has enough stuff so I won't get lost in it.''} Besides, P3 found it very convenient not to describe the video content to \system: \textit{``I always need to describe a specific situation before asking how to do it [with ChatGPT].''} Given the inconvenience of practicing directly with ChatGPT, it is necessary to develop a tool that could provide video-specific guidance like \system.

\subsubsection{\system reveals the potential for personalized learning.} Among participants who do not practice while watching, a common reason is ``Uncertain about how to start or what to practice''. With \system, learners could learn the video content guided by learning goals. According to P8: \textit{``it's better if I can choose learning goals I wanna learn.''} Another participant (P14) pointed out, \textit{``I dislike reading texts about code''.} P9 said, \textit{``I prefer writing code than answering questions.''} Using our action configurator, we can flexibly support these learner preferences. Participants also proposed additional features that could be added to make the code cells more integrated with \system, for example, P16: \textit{``Add code function explanation next to each line of the code instead of in the chat interface''}.

\subsubsection{Cognitive load of \system's interface.} \system presented a moderate level of intrinsic load, minimized extraneous cognitive demands, and maximized germane cognitive engagement for learning. In the cognitive load survey, we use a 10-point scale where 0 represents ``not at all the case'' and 10 represents ``completely the case''. The Intrinsic Load (IL), which represents task complexity and the learner's prior knowledge, averaged at 5.08 ($SD = 0.42$). This score was in the middle of our scale, indicating that participants found the task complexity to be moderately challenging. Extraneous Load (EL) reflects unnecessary cognitive load imposed by instructional features, with a mean score of 2.75 ($SD = 0.66$). This significantly lower score indicates that instructional elements are largely perceived as not unduly adding to cognitive load, suggesting that they are carefully designed to avoid unnecessary complexity.  Germane Load (GL), associated with cognitive processes beneficial for learning, stood out with a mean of 8.05 ($SD = 0.37$). This indicates that the vast majority of participants found the instructional features to significantly enhance their learning and understanding, suggesting an effective instructional design aimed at promoting meaningful cognitive processing. The relatively low standard deviations indicate a general consensus among participants regarding this assessment.

\subsubsection{Overall usability of \system's interface.} In general, participants responded positively to the learning process with \system. In the PSSUQ survey, we anchored at the endpoints with the terms ``Strongly agree'' for 7 and ``Strongly disagree'' for 1. We found the overall usability score for the system was 6.16 ($SD = 1.27$). In terms of sub-scales, the System Usefulness (SYSUSE) score was 6.25 ($SD = 1.23$), indicating that the system was perceived as relatively easy to use and learn by the respondents. The Information Quality (INFOQUAL) score is 5.95 ($SD = 1.43$), suggesting that the information provided by the system was generally clear and helpful, but that there may be room for improvement in specific areas such as line-by-line code explanation and debugging assistance. The Interface Quality (INTERQUAL) score is 6.28 ($SD = 1.06$), which means that participants found the system’s interface to be reasonably pleasant and visually appealing. These results suggest a good level of \system's usability overall but highlight potential areas for enhancement, especially in making information more accessible and understandable to further improve the user experience.

\subsection{\system's Failure Modes}
Our user study also highlighted a range of failure patterns that the videos, the LLMs, or the system itself could cause. For instance, at times, the video clips corresponding to the learning goals are too short or too long, which affects the learning experience. We noticed that because video creators solved problems of varying difficulty at different speeds, a video clip can be as short as 12 seconds or as long as 348 seconds. One participant (P15) noted that it would be better if she could watch the video more coherently: \textit{``Just from my point of view, the Tutorials sometimes were cut a little suddenly. So, for people who don't know data analysis so much, it's a challenge for them to give a reaction immediately.''} Similarly, one participant (P4) expressed that it is a waste of time to watch the video author correct his own errors (common in unedited screencasts). Merging adjacent short video clips or removing unnecessary parts from long video clips is a venue for future work. Additionally, waiting for LLMs to generate a message can be frustrating. Several participants observed that waiting for an incoming message from \system takes too long. According to our observations, the waiting time is generally between ten seconds to fifteen seconds for one message, which considerably slows down learning efficiency. When we use a self-built chatbot framework instead of the pre-built implementation by LangChain, the waiting time is reduced to about five seconds, which appears acceptable. However, in order to achieve faster response speed, LLMs should be iterated out for faster models with small to moderate context window to cope with chat mode.
\section{Discussion}
In \system, we employ the \framework framework to realize a conversational LLM tutor system to assist in learning from programming videos. We implement a novel pipeline and DSL to generate high-quality conversations while allowing domain and content experts to specific key pedagogical strategies. Our user study findings highlight the benefits, limitations, and opportunities of using \system. Here, we elaborate on these limitations and discuss future opportunities for improvements. 

\subsection{Broader Utility}
Beyond a mentorship learning context, a compelling application of \system is in pair programming. Pair programming is a technique in which two individuals share a single computer as they work together to develop software. In traditional pair programming, one person actively writes code (the ``driver''), and the other person provides guidance, feedback, and suggestions (the ``navigator''). We can imagine our \systems prompting techniques can be extended to support collaborative brainstorming and problem-solving in open-ended domains. Rather than relying on video and transcript, the system might build on code examples and datasets to generate productive interactions. Such approaches are shown to provide a variety of benefits, including increased code quality, productivity, creativity, knowledge management, and self-efficacy \cite{10.1145/3159450.3159516, 10.1145/3017680.3017748}.

Second, we imagine content creators can leverage \system to provide more engaging viewing and learning experiences within their own platforms to scale the utility of their content. As domain and topic experts, they can take a human-in-the-loop approach to configure the prompt behavior and contexts. Unlike straightforward factual information, their expertise involves deep understanding, the ability to navigate nuances based on experiential and tacit knowledge, and the application of knowledge in varied, often unpredictable, real-world situations. \system can be used to formulate novel learning goals and teaching moves to help learners acquire these skills. Third, we imagine \system being useful in formal classrooms. Since the pandemic, many courses have been offered in a hybrid format with recorded lectures for students to watch asynchronously. We imagine \system can be configured by instructors to integrate and interleave watching and completing specific assignments set by the instructors. Further, teaching assistants might be able to personalize the prompt parameters in \system to provide better guidance and practice on hard-to-grasp concepts and for low-achieving students.  Rather than accompanying students to practice their ability to write code, \system can also offset the burden on teaching assistants by helping students to find problems, think about problems, and solve problems while watching videos.

\subsection{Limitations and Future Work}
\subsubsection{Adding Novel Interventions} In \systems conversational interactions, we allow experts to specify well-known intervention strategies such as multiple-choice-questions (MCQs) or fill in the blank. There is an opportunity to support more expressive interactions, such as allowing students to directly annotate over visualizations to engage in diagram construction, etc. However, our current implementation favors generalizability over integrating topic-specific techniques. Future research can investigate ways for LLMs to transform users' natural language description into a domain-specific language for domain-specific interaction formats. Alternately, the DSL can be extended to include techniques for generating interactive widgets that are interpretable for the student model.  

\subsubsection{Cross-Platforms and Cross-Language Learning} We deploy \system as an extension in JupyterLab that provides a convenient platform for interactive data science and scientific computing. However, users may have preferences for an integrated development environment (IDE), such as using R Studio for R, PyCharm for Python, XCode for C++, or Visual Studio Code. Therefore, future work can extend \system to multiple IDEs as plug-ins. A different scenario is when students are learning game development; they not only need IDE to write C\# code but also need Unity to watch and manipulate. This will require cross-platform integration. More importantly, we imagine future iterations of \system can take an input video in one language, such as R, and generate code and instructions in a different language, such as Python while keeping the underlying learning content the same.

\subsubsection{Supporting Applications in Diverse Domains} Although our evaluation study provides valuable insights into \system's usage in exploratory data analysis videos, this topic does not encompass all the programming videos. Based on user feedback, many people also want to learn topics such as machine learning, game development, etc. Our technical evaluation has proven that the current DSL structure can support conversation for other topics. More broadly, how to support not only programming learning but also more everyday life teaching videos, such as cooking, makeup, and fitness, opens up problems for inquiry and solutions in the future. There are already some works that support the learning processes of life-teaching videos \cite{Truong2021AutomaticGO}. However, students may need to be provided with feedback rather than just learning. Multi-modal LLMs may be a potential solution to monitor and analyze student performance in the future.

\section{Conclusion}

We developed \system, a system for assisting students learn programming while watching online tutorial videos. While using \system, learners interact with an LLM-powered tutor through rich conversational interactions and follow the guidance and feedback to meet their learning goals. \system design is informed by the \framework framework in learning sciences, and backed by a student model to generate targeted pedagogical strategies to support effective learning. In other words, \system dynamically adapts its teaching strategies based on the observed learning progress of students. Our prompting pipeline is generalizable across different topics, actions, and learning goals. Our user evaluation reveals that \system augments video-based learning experiences, is aligned with learning goals, and does not increase the learner's cognitive workload. Using \system, learners can effectively engage in self-directed learning to acquire and apply the concepts and techniques contained in programming videos.






\bibliographystyle{ACM-Reference-Format}
\bibliography{99_refs}

\clearpage
\appendix
\appendix
\section{Appendix}

\subsection{Cognitive Apprenticeship Methods}
\begin{figure*}[t!]
  \centering
  \includegraphics[width= \textwidth]{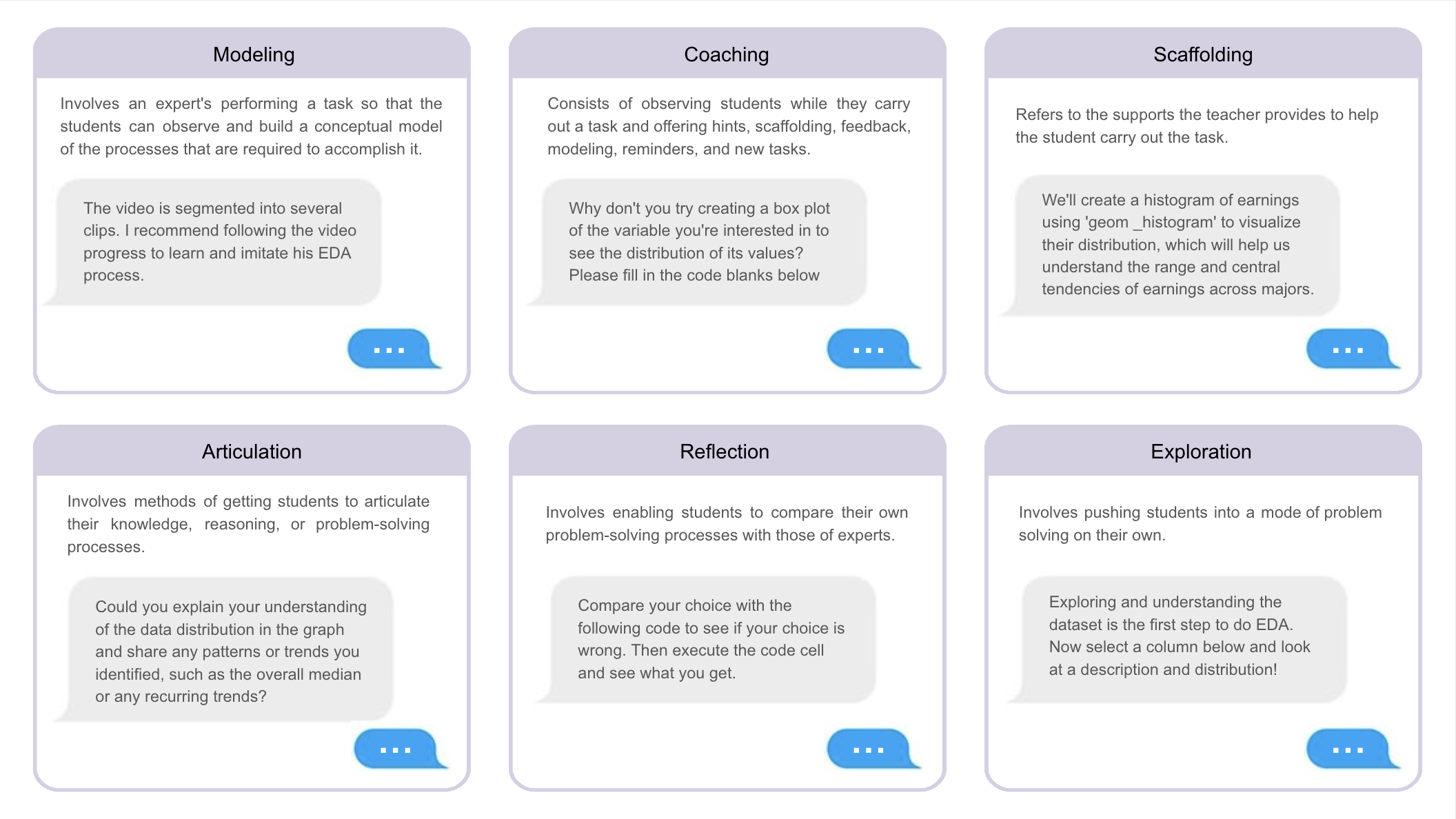}
  \caption{The six cognitive apprenticeship methods and an example message of each method.}
  \label{fig:methods}
\end{figure*}

Figure~\ref{fig:methods} shows the definition of the six cognitive apprenticeship methods and their corresponding messages.

\subsection{Domain Specific Language Structure}
Below is an example of the DSL structure for the video segment ``Visualize the data'' when the student has a median skills level.

\begin{promptbox}
\begin{lstlisting}[language=json, breaklines=true]
[
    {
        "knowledge": "Declarative knowledge: The task is visualizing the distribution of median earnings across major categories using a box plot and enhancing readability by reordering categories and formatting axis labels.",
        "actions": [
            {
                "method": "Scaffolding",
                "action": "Demonstrate the current task and provide explanations of the concepts underlying the current step of the task using plain-text",
                "prompt": "[Use one sentence to explain the Declarative knowledge: The task is visualizing the distribution of median earnings across major categories using a box plot and enhancing readability by reordering categories and formatting axis labels. at this step, such as what effect we want to achieve, why we do it, and what function we use to do it]",
                "interaction": "plain-text",
                "parameters": ["knowledge"],
                "need-response": false
            }
        ]
    },
    {
        "knowledge": "Procedural knowledge: To achieve a clear visualization of categorical data distributions, one must use 'geom_boxplot' on 'ggplot' in R because it effectively displays the spread and central tendency of the data.",
        "actions": [
            {
                "method": "Scaffolding",
                "action": "Demonstrate the current task and provide explanations of the concepts underlying the current step of the task using plain-text.",
                "prompt": "[Use one sentence to explain the Procedural knowledge: To achieve a clear visualization of categorical data distributions, one must use 'geom_boxplot' on 'ggplot' in R because it effectively displays the spread and central tendency of the data. at this step, such as what effect we want to achieve, why we do it, and what function we use to do it]",
                "interaction": "plain-text",
                "parameters": ["knowledge"],
                "need-response": false
            },
            {
                "method": "Coaching",
                "action": "Use fill-in-blanks to guide the student through practice exercises, offering targeted hints and feedback.",
                "prompt": "[Use one sentence to prompt the student to fill in the {code-line-with-blanks} below][Provide a brief hint to help them through it]",
                "interaction": "fill-in-blanks",
                "parameters": ["code-line-with-blanks"],
                "need-response": true
            }
        ]
    },
    {
        "knowledge": "Procedural knowledge: To achieve an ordered factor level based on the 'Median', one must use 'fct_reorder' on 'Major_category', because it facilitates easier comparison across categories by sorting them from lowest to highest median earnings.",
        "actions": [
            {
                "method": "Scaffolding",
                "action": "Demonstrate the current task and provide explanations of the concepts underlying the current step of the task using plain-text.",
                "prompt": "[Use one sentence to explain the Procedural knowledge: To achieve an ordered factor level based on the 'Median', one must use 'fct_reorder' on 'Major_category', because it facilitates easier comparison across categories by sorting them from lowest to highest median earnings. at this step, such as what effect we want to achieve, why we do it, and what function we use to do it]",
                "interaction": "plain-text",
                "parameters": ["knowledge"],
                "need-response": false
            }
        ]
    },
    {
        "knowledge": "Procedural knowledge: To achieve improved readability of axis labels, one must use 'coord_flip' and 'scale_y_continuous' with 'dollar_format' on the plot because flipping the coordinates helps in reading long category names and dollar formatting makes the earnings data more interpretable.",
        "actions": [
            {
                "method": "Scaffolding",
                "action": "Demonstrate the current task and provide explanations of the concepts underlying the current step of the task using plain-text.",
                "prompt": "[Use one sentence to explain the Procedural knowledge: To achieve improved readability of axis labels, one must use 'coord_flip' and 'scale_y_continuous' with 'dollar_format' on the plot because flipping the coordinates helps in reading long category names and dollar formatting makes the earnings data more interpretable. at this step, such as what effect we want to achieve, why we do it, and what function we use to do it]",
                "interaction": "plain-text",
                "parameters": ["knowledge"],
                "need-response": false
            },
            {
                "method": "Reflection",
                "action": "Encourage students to review and debug their code using show-code, and to reflect on the learning process by executing the complete code block to verify their understanding.",
                "prompt": "[Use one sentence to let the student compare his answer with the standard {code-block}][Use one sentence to encourage the student to execute the complete code block to verify his understanding]",
                "interaction": "show-code",
                "parameters": ["code-block"],
                "need-response": true
            }
        ]
    }
]
\end{lstlisting}
\end{promptbox}

\subsection{Action Set}
Table~\ref{tab:cognitive_moves} shows the action set that generates prompts.

\begin{table*}[ht!]
\centering
\renewcommand{\arraystretch}{1.5}
\begin{tabular}{p{2cm}|p{6cm}p{6cm}}
\hline
\textbf{Move} & \textbf{Action} & \textbf{Prompt} \\ \hline
\multicolumn{3}{c}{\textit{Programming-Related}} \\ \hline
Scaffolding & Demonstrate the current task and provide explanations of the concepts underlying the current step of the task using \textbf{plain-text}. & [Use one sentence to explain the \{knowledge\} at this step, such as what effect to achieve, why we do it, and what function we use to do it] \\ \hline
Coaching & Use \textbf{fill-in-blanks} to guide the student through practice exercises, offering targeted hints and feedback. & [Use one sentence to prompt the student to fill in the \{code-line-with-blanks\} below to practice the \{knowledge\}][Provide a brief hint to help them through it] \\ \hline
Articulation & Use \textbf{plain-text} to allow students to articulate their understanding of knowledge. & [Use one sentence to ask the student to explain their understanding and reasoning about \{knowledge\}, such as articulate why make this kind of visualization rather than others] \\ \hline
Reflection & Encourage students to review and debug their code using \textbf{show-code} and to reflect on the learning process by executing the complete code block to verify their understanding. & [Use one sentence to let the student compare their answer with the standard \{code-block\}][Use one sentence to encourage the student to execute the complete code block to verify their understanding] \\ \hline
\multicolumn{3}{c}{\textit{Concept-Related}} \\ \hline
Scaffolding & Provide structured guidance through \textbf{plain-text} as the student works on the task to learn the \{knowledge\}. & [Use no more than three sentences to guide the student step by step on how to learn and apply the \{knowledge\}, the student has made the visualization] \\ \hline
Coaching & Use \textbf{multiple-choice} to observe the student's approach to tasks, offering feedback to guide learning. & Propose a multiple-choice question for the student to understand the \{knowledge\}, such as what could be the potential reason behind the pattern \\ \hline
Articulation & Encourage students to use \textbf{interaction} to verbally explain their thought process and reasoning behind their observations and conclusions. & [Use one sentence to ask the student to explain their understanding and reasoning about \{knowledge\}, such as articulate what patterns they found in the chart] \\ \hline
Reflection & Encourage students to use \textbf{plain-text} to self-evaluate their performance, identifying strengths and areas for improvement. & [Use one sentence to give feedback on the \{student-answer\}][Use one sentence to tell the student if any additional steps could confirm their choice][Ask the student to remember the choice and see if it makes sense as they watch the rest of the video] \\ \hline
\end{tabular}
\caption{Programming and concept-related cognitive apprentice moves in teaching programming and concepts. The bold terms are interactions used in prompts. Parameters are quoted in curly brackets.}
\label{tab:cognitive_moves}
\end{table*}

\subsection{Prompts for LLMs}
Below are the prompts we fed into the ChatGPT API to perform each of \system's key algorithms.

\subsubsection{Video Segmentation}
Below is the prompt doing \textbf{Summarize} in the video segmentation algorithm.
\begin{promptbox}
\textbf{Summarize:} Here is a video transcript about \{video\_topic\}. Summarize the video content that corresponds to each given learning goal.
The transcript is not necessarily arranged in the order in which the learning goals are defined and can contain multiple segments with the same learning goal.
The script may contain only some of the learning goals. Please do not include summary of learning goals that do not exist in the transcript.
Increase the granularity. For example, if the video author creates two different visualizations, they should be summarized into two points.

Response only in a list in the order of their appearance in the video without any explanations, for example:
\begin{verbatim}
[
    ("Introduction", summary),
    ("Load data/packages", summary),
    ("Understand the dataset", summary),
    ("Visualize the data", summary),
    ("Interpret the chart", summary),
    ("Visualize the data", summary),
    ("Interpret the chart", summary),
    ("Preprocess the data", summary),
    ...
]
\end{verbatim}
\end{promptbox}

\subsubsection{Knowledge Summarize}
Below is the prompt summarizing the declarative and procedural knowledge in a concept-related video segment. The prompts used for the programming-related clips differed only in the definition of the knowledge structure.

\begin{promptbox}
The following \{video\_type\} video transcript is about a learning goal: \{learning\_goal\}. Summarize the declarative and procedural knowledge in the video transcript.

The result should be summarized in one sentence of procedural knowledge and no more than \{num \- 1\} sentences of declarative knowledge in the order in which it should be learned.

Each piece of knowledge should follow this format:

Procedural knowledge: ``To achieve/understand + [specific goal/outcome] + one need to + [general actions/processes] + [additional details] + and consider/use + [relevant factors/tools].'' The [general actions/processes] should be quoted in a pair of \& sign.

For example, ``To understand the distribution of earnings by college major, one need to \&examine the histogram and identify overall trend or extreme values\&, and consider whether high earnings are due to the field's financial reward or influenced by factors such as low sample size and high variation.''

Declarative knowledge: ``[Subject] + [verb phrase] + that + [independent clause]''.

For example, ``The median income by college major shows that majors earn a median income of over \$30K right out of college.''
And sort the output knowledge order according to the correct cognitive order. For example, students need to first learn how to interpret the chart then find out the facts in the chart.

Your response should be in a list format without any explanations:
\begin{verbatim}
[
    'knowledge_1',
    'knowledge_2',
    ...
]
\end{verbatim}
\end{promptbox}

\subsubsection{Teaching Methods Arrangement}
Below is the prompt for arranging the teaching methods according to the cognitive apprenticeship framework.

\begin{promptbox}
You are an expert mentor who is good at arranging teaching methods to help students learn from a video about \{video\_type\}.
You are teaching students to learn knowledge for \{learning\_obj\} using the Cognitive Apprenticeship framework.
Definition of Cognitive Apprenticeship framework methods:

Coaching: mentor observes mentee's activities along with provision of guidance and feedback

Scaffolding: mentor supports mentee while they work through the task with gradual fading of such supports

Articulation: mentor encourage mentees to verbalize their knowledge and thinking

Reflection: mentor enable mentees to self-assesses own performance

Your task is to choose proper Cognitive Apprenticeship methods to teach the student the given knowledge.
The input knowledge list contains the declarative and procedural knowledge in the video.
The input student mastery level list has a one-to-one correspondence with the knowledge, representing the student's mastery of each knowledge.

Teaching method arrangement rules:

1. Global before local skills: use Scaffolding as the first move to teach the first knowledge.

2.1 Increasing complexity for concept-related video:
Choose one method from Scaffolding, Coaching, or Articulation to teach each knowledge.
If the student's mastery level of the corresponding knowledge exceeds 0.5, Scaffolding should fade out.
Coaching should be followed by Reflection. Reflection should only be used after Coaching.

2.2 Increasing complexity for programming-related video:
For each declarative knowledge, choose between Scaffolding and Articulation.
For each procedural knowledge, if the student's mastery level of the corresponding knowledge is lower than 0.3, use Scaffolding only.
If the student's mastery level of the corresponding knowledge is between 0.3 and 0.7, use Scaffolding and Coaching.
If the student's mastery level of the corresponding knowledge is higher than 0.7, Scaffolding fades out, and Coaching is used only.
Reflection should be used once as the last method for the last knowledge.

3. Increasing diversity: diversify the selection of teaching methods based on the first two conditions.

You should use no more than three methods for each knowledge. Please include your choice in a structure in the same format like the following.
Response Example:

\begin{verbatim}
[
    {{
        "knowledge": ...,
        "method": [...]
    }},
    ...
]
\end{verbatim}
\end{promptbox}

\subsubsection{Conversation}
The prompt used for generating dialog messages is as follows:
\begin{promptbox}
You are an expert in Data Science, specializing in \{video\_type\}. Your task is to use the Cognitive Apprenticeship approach to assist a student in learning \{video\_type\} through David Robinson's Tidy Tuesday tutorial series.

You will be provided with one or more of the following inputs:

- knowledge: the knowledge that will be learned by the student

- pedagogy: the specific cognitive apprenticeship move you must follow to guide students.

- student's code or question or choice: the student's current performance, encompassing either the code in the student's notebook or the student's query sent to you or the student's choice in the multiple-choice question.

- other parameters or requirements: additional information or requirements you must follow to guide the student.

Notes for Response:

- Don't answer or say anything irrelevant to the topic (\{video\_type\}) or the programming language (\{kernel\_type\}).

- Use natural language to communicate in the first person as a teaching assistant.

- You must strictly follow the pedagogy to provide guidance.

- Tailor your advice to the programming language the student uses: \{kernel\_type\}.

- Don't tell the student your response is based on the transcript or code.

- You can find out the full list of conversation history below.

\end{promptbox}

\label{sec:appendix}

\end{document}